\documentclass{aa}  

\usepackage{graphicx}
\usepackage{txfonts}
\usepackage{upgreek}
\usepackage{ulem}
\usepackage{arydshln}
\usepackage{multicol}
\usepackage{placeins}
\usepackage{lastpage}

\newcounter{FootnoteA}
\newcounter{FootnoteB}

\begin{document} 
    \title{Enhanced spatial resolution and performance for $N$-body simulation of self-interacting dark matter}
    \titlerunning{Enhanced spatial resolution SIDM simulations}

   \author{Moritz S.\ Fischer
          }
    \authorrunning{Fischer}

    \institute{
        Donostia International Physics Center (DIPC), Paseo Manuel de Lardizabal 4, 20018 Donostia-San Sebastián, Spain\label{inst:dipc}\\
        \email{moritz.fischer@dipc.org}
    }

   \date{Received XX Month, 2026 / Accepted XX Month, 20XX}

  \abstract{Studying the implications of self-interacting dark matter (SIDM) on astrophysical objects often requires simulations. $N$-body simulations are a standard tool but face a few challenges. One is resolving the mean free path in gravothermally collapsing halos. This is necessary to obtain accurate results and is achieved in state-of-the-art simulations by increasing the number of resolution elements.}
  {I aim to formulate a numerical scheme that gives an enhanced spatial resolution compared to those methods and allows increasing the resolution without changing the number of resolution elements.}
  {The novel scheme that I present is formulated by discarding the classic approach of using kernel functions and computing interaction probabilities. Instead, pairs of particles always interact if their distance is below a specific threshold. Importantly, this distance threshold depends on their relative velocity and the size of the time step. This new formulation no longer requires some of the computations in state-of-the-art schemes and thus helps to increase the performance.}
  {Through a variety of test problems with known solutions, I study the accuracy of the new scheme and demonstrate its advantages over state-of-the-art SIDM formulations. Notably, the scheme reduced the computational costs, making the SIDM computations typically much cheaper than gravity. As a consequence, a typical SIDM simulation of isotropic scattering requires only a few percent of its total computational cost for the self-interactions.}
  {In conclusion, the new scheme presents an improvement over previous schemes and helps to address current challenges. Besides the enhanced performance, it allows us to achieve a higher accuracy when simulating gravothermally collapsing SIDM halos.}
  \keywords{methods: numerical -- dark matter}

   \maketitle

\section{Introduction} \label{sec:introduction}

Self-interacting dark matter (SIDM) is a class of particle physics models for dark matter (DM) that allows the particles to interact with each other in a non-gravitational way. 
This is in contrast to the cosmological standard model, Lambda Cold Dark Matter ($\Lambda$CDM), where DM is assumed to interact only gravitationally; other interactions, if present, have a negligible effect.
SIDM was proposed by~\cite{Spergel_2000} to address historical discrepancies between DM-only simulations and observations.\footnote{There is even earlier work by \cite{Carlson_1992} that considered DM self-interactions in a cosmological context.}
The non-gravitational scattering of SIDM holds the potential to explain tensions and discrepancies between observations of astrophysical small-scale structure, such as galaxies, and predictions of $\Lambda$CDM~\citep[e.g.][]{Tulin_2018, Adhikari_2025, Nadler_2026}.
Of particular interest are gravothermally collapsing SIDM halos, where the self-interactions give rise to an increasing central DM density by transporting energy outward. These halos could account for observations that hint at compact dark objects~\citep[e.g.][]{Minor_2020, Yang_2021, Nadler_2023, Ragagnin_2024, Li_2025, Zhang_2025, Fischer_2026a, Kong_2026, Yu_2026, Gilman_2026b}.

For the astrophysically interesting cross-sections, SIDM behaves as mildly collisional. As a consequence, it cannot be approximated as collisionless or treated as a fluid.
Instead, the Boltzmann equation together with self-gravity needs to be solved to describe the evolution of SIDM.
Several methods have been developed to solve the Boltzmann equation numerically, a problem that is relevant for various areas of physics~\citep[e.g.][]{Dimarco_2014, Akhlaghi_2023}.
To study the impact of DM self-interactions on galaxies and other astrophysical objects, N-body simulations are a common method.
Here, a coarse-grained distribution of numerical particles represents the underlying physical DM distribution and evolves under the influence of gravity and self-interactions over time. The first simulation of that kind was undertaken by \cite{Burkert_2000} and substantially improved by \cite{Kochanek_2000}.
Interestingly, a similar approach in the context of Direct Simulation Monte Carlo \citep[DSMC, e.g.][]{Bird_1963, Bird_1994} schemes had been developed by~\cite{Olson_2008}.
As of today, these simulations have been further refined in terms of numerical fidelity and the variety of particle physics models that can be simulated \citep[e.g.][]{Fischer_2026b, Rosenlyst_2026}.

One limitation of the SIDM $N$-body simulations is that the mean free path needs to be resolved, i.e.\ the separation of the scattering particles should roughly not exceed the mean free path. This had been first noted by~\cite{Koda_2011} and later also demonstrated by \cite{Fischer_2024c, Fischer_2025b}. In this paper, I address this issue by deriving a formulation that minimises the interaction distance between the scattering particles and thus helps to resolve the mean free path.

To do so, I deviate from the traditional scheme that computes an interaction probability based on pairs of particles. This probability is replaced by an interaction distance. If the particles are closer than this separation, they interact; otherwise they don't. Thus, I do not draw pseudo-random numbers to decide whether they interact, but instead assume that the distribution of the numerical particles itself is random. The scheme has similarities with the one introduced by~\cite{Huo_2019}. They use a geometric interpretation of the cross-section to compute a distance within which their numerical particles interact. However, this distance is not adaptive; e.g.\ it is independent of the simulation time step and can be disadvantageous for the time stepping of the simulation. In contrast, my new formulation depends on the time step and thus, the time step of the particles can be better adjusted, allowing for an enhanced performance of the simulation.
The traditional scheme that comes the closest to my new scheme in terms of how the interaction probability is computed might be the one by~\cite{Robertson_2017a}, as their interaction probability is independent of the separation of the particles as long as their separation does not exceed the kernel size. In detail, they use a one-sided top hat kernel.
This is similar because within the interaction radius, I assume a constant interaction probability as well, but a larger one, namely unity. However, their formulation does not choose the kernel size in a way to minimise the separation of interacting particles; instead, it is either constant or adapted to the local density as done in many SIDM implementations~\citep [e.g.][]{Robles_2017, Banerjee_2020, Yang_2020, Fischer_2021a, Ebisu_2022, Valdarnini_2024, Zier_2026}.
In summary, one may view the new scheme that I present here to some extent as a combination and generalisation of the work by \cite{Robertson_2017a} and \cite{Huo_2019} with the addition of an adaptive interaction distance that depends on the relative velocity and the time step size. At the same time, the other advantages of the SIDM formulation as presented by \cite{Fischer_2026b}, such as the explicit energy and momentum conservation, are kept.

In addition to formulating this novel scheme, I demonstrate that it successfully reproduces test problems with known solutions. Furthermore, I simulate a gravothermally collapsing SIDM halo and highlight the advantages of the new scheme for the deep collapse phase.
The new scheme allows us to increase the spatial resolution for DM self-interactions and to speed up the $N$-body simulations. No shortcomings are found compared to commonly used methods for modelling DM self-interactions. Only the assumption that the particle positions need to be sufficiently random needs to be fulfilled.
Lastly, I note that I implemented the novel scheme in the cosmological simulation code \textsc{OpenGadget3} \citep{Dolag_2026}. It is a successor of \textsc{gadget-2} \citep{Springel_2005} that has undergone several extensions and improvements \citep[e.g.][]{Ragagnin_2016}.\footnote{The code can be retrieved from the public \textsc{OpenGadget3} repository: \url{https://gitlab.lrz.de/AstroCodes/OpenGadget3}} The implementation will become public in the near future.

The remainder of the paper is structured as follows. In Sect.~\ref{sec:methods}, I derive the new formulation for SIDM in $N$-body simulations. Various tests of the new scheme follow in Sect.~\ref{sec:tests} to examine whether it reproduces specific test problems with known solutions, and in Sect.~\ref{sec:scaling} to perform strong scaling tests. In Sect.~\ref{sec:halo} I illustrate the advantages of simulating the gravothermal evolution of an isolated halo. I discuss the properties of the scheme and further directions in Sect.~\ref{sec:discussion}. Finally, in Sect.~\ref{sec:conclusion}, I summarise and conclude.

\section{Formulation of the scheme} \label{sec:methods}

In this section, I derive the formulation of the new scheme, which aims to minimise the distance between interacting particles.
I do this in the context of large-angle scattering (see Sect.~\ref{sec:methods_largeangle}), i.e.\ where the scattering rate of the particles is sufficiently low, so that one can afford to resolve the individual scattering events in time~\citep[e.g.][]{Burkert_2000, Kochanek_2000}. Next, a derivation for the frequent SIDM scheme~\citep{Fischer_2021a} follows (see Sect.~\ref{sec:methods_smallangle}). This scheme is designed for very anisotropic cross-sections~\citep{Kahlhoefer_2014}, i.e.\ small-angle scattering, where the scattering is too frequent to resolve individual scattering events.\footnote{Although the scattering can be very frequent, it can have a small effect on the DM distribution, because an individual scattering event has a tiny effect as the interacting particles are only slightly deflected. As a consequence, very anisotropic cross-sections can give rise to extreme scattering rates while staying within current bounds on the SIDM cross-section.}
Following these derivations, I discuss the neighbour search in the context of this new scheme (Sect.~\ref{sec:methods_neighboursearch}) and elaborate on the limitations arising from successive interactions and mitigation strategies (Sect.~\ref{sec:methods_successive_interactions}).
I end by discussing the formulation of an SIDM time step criterion (Sect.~\ref{sec:methods_timestep}).

\subsection{Large-angle scattering} \label{sec:methods_largeangle}

A central aim of the new scheme is to minimise the distance between the interacting particles. In most $N$-body SIDM schemes, particles within a given distance from each other, typically related to a chosen kernel size, can interact. In practice, a random set of particles within that distance interacts, but not necessarily the closest ones. But if one instead would only choose the closest\footnote{Strictly speaking, it is more complicated than choosing the closest ones, as also the velocity of the particle and the velocity dependence of the cross-section plays a role. This becomes clear with Sect.~\ref{sec:methods_largeangle}.} particles, the distance, $|\textbf{x}_i - \textbf{x}_j|$, between the interacting particles would be minimised, leading to a better spatial resolution. However, there is also a caveat as I discuss in Sect.~\ref{sec:methods_successive_interactions}.

As I aim to minimise the interaction distance, I make an important assumption about the distribution of numerical particles. Namely, that their positions are random, i.e.\ one can interpret their location as randomly drawn from the actual physical density distribution. This means that finding two particles closer than a specific distance can be associated with a probability.

The traditional interaction probability assuming a top hat kernel with kernel size $h$ can be expressed as~\citep[see Eq.~6 by][]{Robertson_2017a}
\begin{equation} \label{eq:interaction_probability}
P_{ij} = \frac{\sigma}{m_i} \, \Delta t \, |\Delta \mathbf{v}_{ij}| \, m_{\mathrm{n},i} \left(\frac{4}{3} \, \uppi \, h \right)^{-1} \,.
\end{equation}
Here, $\sigma / m_i$ is the total cross-section per physical DM particle mass. Using the scattering angle $\theta_\mathrm{cms}$ in the centre-of-mass frame, the total  cross-section is given by
\begin{equation} \label{eq:total_cross_section}
\sigma=\int_{-1}^{1} \frac{\mathrm{d} \sigma}{\mathrm{d} \cos \theta_{\mathrm{cms}}} \, \mathrm{d} \cos \theta_{\mathrm{cms}}
\end{equation}
and proportional to the scattering rate.
In Eq.~\eqref{eq:interaction_probability}, the last factor is the mass $m_{\mathrm{n},i}$ represented by the numerical particle $i$ divided by the volume of a sphere with radius $h$. This gives us the density within the top-hat kernel. Imagining that a physical DM particle is moving through this constant density with a relative velocity, $\Delta \mathbf{v}_{ij}$ over a time span $\Delta t$, one obtains the probability $P_{ij}$ that it scatters. Strictly speaking, this is the expectation value of how often a particle scatters within $\Delta t$, i.e.\ the scattering rate $R = P_{ij} / \Delta t$ \citep[see also sect.~4 of][]{Fischer_2024b}.

The interaction radius $r_{ij}$ for the new scheme can be obtained by setting $P_{ij} = 1$ and identifying $h = r_{ij}$. This means that one chooses the kernel size so small that the interaction probability equals unity. Then the kernel size gives us the interaction criterion, i.e.\ the distance below which all particles should interact, while above no interaction takes place.
Hence the interaction radius is
\begin{equation} \label{eq:interaction_radius_rare}
r_{ij}|_\mathrm{rare} = \sqrt[3]{\frac{3}{4 \uppi} \frac{\sigma(|\Delta \mathbf{v}_{ij}|)}{m_i} \, m_{\mathrm{n},i} \, \Delta t \, |\Delta \mathbf{v}_{ij}}| \,.
\end{equation}
As I assume that the particle positions are random, I can select the `random’ interaction partners based on the distance criterion $|\textbf{x}_i - \textbf{x}_j| < r_{ij}|_\mathrm{rare}$, instead of drawing pseudo-random numbers as traditionally done.
Particles that fulfil this criterion interact, i.e.\ the direction of their velocity vectors in the centre-of-mass frame is chosen randomly according to the differential cross-section using pseudo-random numbers. In detail, the procedure is identical to the one described by \cite{Fischer_2026b}.

\subsection{Small-angle scattering} \label{sec:methods_smallangle}

In addition to the scheme for large-angle scattering, I derive a new formulation for the scheme simulating the limit of a very anisotropic cross-section. In contrast to computing an interaction probability, as is done for large-angle scattering, small-angle scattering is simulated by computing a drag force and applying perpendicular momentum diffusion~\citep{Fischer_2021a}. Here, all particles are deflected by a small angle if they are within a distance related to the chosen kernel size. This is analogous to the scattering probability for large-angle scattering, where not all particle pairs within that distance have the maximum probability, i.e.\ unity. 
Similarly, not all particle pairs for small-angle scattering experience the maximal drag force. Importantly, there is not a clear maximal drag force, but the larger the corresponding deflection angle, the larger the numerical errors. Hence, given an acceptable error, one can choose a maximum deflection angle. In practice, I make this decision in terms of
\begin{equation} \label{eq:tau}
\delta \equiv 2 \frac{|\Delta \mathbf{p}_\mathrm{drag}|}{|\Delta \mathbf{p}_{ij}|} \, .
\end{equation}
Here, $\Delta \mathbf{p}_\mathrm{drag}$ is the momentum change due to the drag force acting parallel to the relative velocity, $\Delta \mathbf{v}_{ij}$, of the two particles. The momentum difference is, $\Delta \mathbf{p}_{ij} = \mathbf{p}_i - \mathbf{p}_j$.
With Eq.~\eqref{eq:tau} the deflection angle in the centre-of-mass frame, $\alpha_\mathrm{cms}$, is fixed via $\delta = 1 - \cos\alpha_\mathrm{cms}$.

Based on a chosen value for $\delta$, one can minimise the interaction distance similar to the large-angle scattering (Sect.~\ref{sec:methods_largeangle}).
The corresponding interaction radius is
\begin{equation} \label{eq:interaction_radius_freq}
r_{ij}|_\mathrm{freq} = \sqrt[3]{\frac{3}{4 \uppi} \frac{\sigma_\mathrm{\Tilde{T}}(|\Delta \mathbf{v}_{ij}|)}{m_i} \, \frac{m_{\mathrm{n},i} \, \Delta t}{\delta} \, |\Delta \mathbf{v}_{ij}}| \,.
\end{equation}
Here, $\sigma_\mathrm{\Tilde{T}}$ denotes the modified momentum transfer cross-section. It is given by
\begin{equation} \label{eq:mod_momentum_transfer_cross_section}
\sigma_\mathrm{\Tilde{T}}= \int_{-1}^{1} \frac{\mathrm{d} \sigma}{\mathrm{d} \cos \theta_{\mathrm{cms}}}\left(1-|\cos \theta_{\mathrm{cms}}| \right) \mathrm{d} \cos \theta_{\mathrm{cms}} \, .
\end{equation}

Given this scheme, all particle pairs with a separation of $ | \textbf{x}_i - \textbf{x}_j| < r_{ij}|_\mathrm{freq}$ experience a drag force-induced parallel velocity change of $|\Delta \mathbf{p}_\mathrm{drag}| = \delta \, |\Delta \mathbf{p}_{ij}| / 2$. From this directly follows the corresponding perpendicular velocity change to model the perpendicular momentum diffusion~\citep[see sect.~2.2 by][]{Fischer_2021a}. The parallel and perpendicular velocity changes together give rise to a deflection by an angle $\alpha_\mathrm{cms} = \arccos(1-\delta)$.
Hence, to model small-angle scattering, all pairs of numerical particles that interact are `scattered' by the same fixed angle $\alpha_\mathrm{cms}$.

\subsection{Neighbour search criterion} \label{sec:methods_neighboursearch}
To find neighbouring particles that could interact, one needs a search criterion, i.e.\ a distance for which one checks if particles could interact.
Here, I employ an upper estimate for the interaction criteria of Eqs.~\eqref{eq:interaction_radius_rare} and~\eqref{eq:interaction_radius_freq} and only check the full criterion for pairs that pass this first check.
The criterion to use when walking the tree is,
\begin{equation} \label{eq:search_radius_rare}
    r_\mathrm{max}|_\mathrm{rare} = \sqrt[3]{\frac{3}{4 \uppi} \, \omega_\mathrm{ext}|_\mathrm{rare} \, m_{\mathrm{n},i} \, \Delta t} \,.
\end{equation}
The factor $\omega_\mathrm{ext}|_\mathrm{rare}$ is a maximum estimate to ensure that no relevant neighbouring particles are missed. It is given by
\begin{equation} \label{eq:omega_rare}
    \omega_\mathrm{ext}|_\mathrm{rare} = \max_{v \, \in \, [0, \,  v_\mathrm{max}]}\left(\frac{\sigma(v)}{m_i} \, v \right) \, .
\end{equation}
Importantly, the index $i$ in Eqs.~\eqref{eq:search_radius_rare} and~\eqref{eq:omega_rare} refers to the same particle species, which is particularly relevant for cases with species that have unequal particle masses.

For the small-angle scattering scheme, the search radius is given by
\begin{equation} \label{eq:search_radius_freq}
    r_\mathrm{max}|_\mathrm{freq} = \sqrt[3]{\frac{3}{4 \uppi} \, \omega_\mathrm{ext}|_\mathrm{freq} \, \frac{m_{\mathrm{n},i} \, \Delta t}{\delta}} \, .
\end{equation}
Similar to large-angle scattering, one needs to compute $\omega_\mathrm{ext}|_\mathrm{freq}$.
\begin{equation}
    \omega_\mathrm{ext}|_\mathrm{freq} = \max_{v \, \in \, [0, \,  v_\mathrm{max}]}\left(\frac{\sigma_\mathrm{\Tilde{T}}(v)}{m_i} \, v \right) \, .
\end{equation}
The search radii as expressed above require $v_\mathrm{max}$. This is the maximum relative velocity a particle experiences in an interaction. For $v_\mathrm{max}$, I use an estimate only and no precise computation. To obtain the estimate, the tree nodes save the minimum and maximum of each component of the velocities of the particles they contain. Based on those minima and maxima, I estimate $v_\mathrm{max}$. To reduce noise, the tree is walked upward until one arrives at a node that contains at least $N_\mathrm{crit}$ DM particles. This is then the node used to estimate $v_\mathrm{max}$.

Lastly, in the case where small- and large-angle scattering is used, the maximum of Eqs.~\eqref{eq:search_radius_rare} and~\eqref{eq:search_radius_freq} is employed in the search. This ensures that no potential interaction partner is missed.

\subsection{The problem of successive interactions}
\label{sec:methods_successive_interactions}

In the following, I discuss a limitation of the scheme described above and how to mitigate it.
If two particles that interacted do not travel far enough over the next time step, they interact again without having had any interaction with another particle in between. These successive interactions are inconsistent with the physics and underestimate the effect that the self-interactions have.

This problem can be avoided by choosing large enough time steps. For the rare scattering scheme, a minimum time step criterion can be formulated. The corresponding time step requirement can be expressed as
\begin{equation} \label{eq:delta_t_successive_interactions}
    \Delta t > \sqrt{\frac{8}{\uppi} \frac{m_{\mathrm{n},i}}{|\Delta \mathbf{v}_{ij}|^2} \frac{\sigma(|\Delta \mathbf{v}_{ij}|)}{m_i}} \, .
\end{equation}
However, in practice this may not be very useful to avoid the problem of successive interactions, as other time step criteria such as the one for gravity may require smaller time steps such that Eq.~\eqref{eq:delta_t_successive_interactions} can no longer be fulfilled.

Instead of increasing the time step, one can increase the search radius such that multiple neighbours are found and compute a probability to interact with them. This makes it much less likely that a successive interaction occurs. 
The default search radius as explained in Sect.~\ref{sec:methods_neighboursearch} for a particle is $r_\mathrm{max} = \max(r_\mathrm{max}|_\mathrm{rare}, r_\mathrm{max}|_\mathrm{freq})$.
The following condition is evaluated, 
\begin{equation} \label{eq:enlarge_crit}
    r_\mathrm{max} > v_\mathrm{max} \, \Delta t \,.
\end{equation}
If it is fulfilled, the search radius is increased as successive interactions are likely to occur.
In detail, $r_\mathrm{enlarge}$, an estimate for the radius that encloses $N_\mathrm{enlarge}$ particles, is computed.
\begin{equation} \label{eq:r_enlarge}
    r_\mathrm{enlarge} = \sqrt[3]{\frac{3}{4\uppi}\frac{N_\mathrm{enlarge} \, m_\mathrm{n}}{\rho}} \,.
\end{equation}
Here, $\rho$ is an estimate for the local DM density using the tree similar to the estimate for $v_\mathrm{max}$ (see Sect.~\ref{sec:methods_neighboursearch}). Therefore, each node keeps track of its DM mass.
The new search radius follows as
\begin{equation} \label{eq:search_radius}
    r_\mathrm{search} = \max(r_\mathrm{max}, r_\mathrm{enlarge}) \,.
\end{equation}

This procedure is only an approximate criterion. In principle, the optimal search radius to avoid the problem of successive interactions may not be independent of the angular- and velocity-dependence of the cross-section. A more sophisticated criterion than evaluating Eq.~\eqref{eq:enlarge_crit} and using Eq.~\eqref{eq:r_enlarge} may allow reducing the number of neighbours evaluated and thus increasing the performance. A detailed study of this is left for future work.

When enlarging the search radius, an interaction probability for large-angle scattering is computed similar to state-of-the-art schemes.
The probability follows from the actual used search radii and the interaction distance (Eqs.~\eqref{eq:interaction_radius_rare} and~\eqref{eq:interaction_radius_freq}).
\begin{equation}
    P_{ij} = \left(\frac{r_{ij}}{\max(r_{\mathrm{search},i}, r_{\mathrm{search},j})}\right)^3 
\end{equation}
In this case, the scheme behaves similarly to the one by \cite{Robertson_2017a}, when employing a top hat kernel with an adaptive kernel size.
In the case of small-angle scattering, the drag force is reduced when enlarging the search radius.
\begin{equation}
    |\Delta \mathbf{p}_\mathrm{drag}| = \delta \, \frac{|\Delta \mathbf{p}_{ij}|}{2} \, \left(\frac{r_{ij}}{\max(r_{\mathrm{search},i}, r_{\mathrm{search},j})}\right)^3 \, .
\end{equation}
This is similar to reducing $\delta$, which also helps to avoid problems with successive scatterings, but is only applied locally.

When the distances are enlarged, the spatial resolution becomes coarser. This can potentially impact the accuracy of the simulation, for example, when it leads to an unresolved mean free path of the scattering. Hence, it is favourable to enlarge the interaction radius only as little as necessary. However, in general, the problem of successive interactions implies a spatial resolution limit.

For simulations with large-angle scattering, one can check whether there are successive interactions happening. Each simulation particle has a unique identification number, ID. For every DM particle, one can save the ID of its last scattering partner; when it interacts again, one can check if the interaction partner is the same as before or a new one. This allows us to estimate on-the-fly how problematic successive interactions are.

For small-angle scattering, successive interactions are less problematic, as the scattering effect in the limit of a small scattering angle can be considered cumulative. However, if a given pair does not only interact twice without undergoing an interaction with any other particle but many times, this can be a problem too. In particular, if the resulting deflection angle becomes large.
Based on the relative velocity of two particles and the time step size, one can estimate how many interactions they would undergo and how large the resulting effect is. This gives rise to a cumulative deflection angle
\begin{equation} \label{eq:cumulative_deflection_angle}
    \alpha_\mathrm{cum} = \frac{4 \, r_{ij} \, \delta}{|\Delta \mathbf{v}_{ij}| \, \Delta t} = 4 \, \left( \frac{\delta}{|\Delta \mathbf{v}_{ij}| \, \Delta t} \right)^{2/3} \, \sqrt[3]{\frac{3}{4 \uppi} \, \frac{\sigma_\mathrm{\Tilde{T}}(|\Delta\mathbf{v}_{ij}|)}{m} \, m_\mathrm{n}}
\end{equation}
that is evaluated on the fly.
The formulation as given by Eq.~\eqref{eq:cumulative_deflection_angle} applies only in the case that the interaction distance is not enlarged.
A large value for $\alpha_\mathrm{cum}$ would indicate that the interaction distance is too small and the interaction distance should be enlarged as described above.
However, this criterion is only an approximation and may not work for all problems.

Lastly, for the discussion of successive interactions, I have assumed equal-mass scattering, while the equations before hold also for unequal-mass scattering. For a discussion on the caveats of unequal-mass scattering, I refer the reader to \cite{Patil_2026}.

\subsection{Time step criterion} \label{sec:methods_timestep}

The aim of a time step criterion for SIDM is to ensure that the time step is sufficiently small to accurately simulate the interactions. In traditional SIDM schemes, it is built to keep the interaction probability sufficiently low or the drag force small enough. However, these aims do not make sense in the context of the scheme discussed here, as the interaction probability cannot exceed unity by construction, and also the strength of the drag force is set by $\delta$.

In the context of the novel scheme, it can make sense to reduce the time step in order to increase the spatial resolution. However, this is only possible within limits, as becomes clear from the explanations about successive interactions given in Sect.~\ref{sec:methods_successive_interactions}.
In many astrophysical situations, the time step implied by the criterion for gravity might be fine, as there is no need to achieve a better spatial resolution.

Importantly, the range where a dedicated SIDM time step criterion could improve the accuracy is limited. It may make sense only in situations where one is not limited by successive interactions. Strictly speaking, this is not only the case when $r_\mathrm{max} \leq v_\mathrm{max} \, \Delta t$ but also when $r_\mathrm{max} \geq r_\mathrm{enlarge}$. A time step criterion could be formulated with the purpose of resolving the mean free path of the scatterings. How large the range is where it could actually fulfil this purpose is an open question. Given the results for the evolution of an isolated halo presented in Sect.~\ref{sec:halo_results}, this doesn't appear to be too important.
As a consequence, I refrain from formulating a time step criterion here and leave this to future work.

Lastly, for the interaction of two numerical particles, it is assumed that the underlying physical DM particles undergo only one scattering event in the large-angle scattering scheme. However, there is always a fraction of physical DM particles that would scatter multiple times \citep[see sect.~4 by][]{Fischer_2024b}. In all $N$-body SIDM schemes, it is assumed that this fraction can be neglected. \cite{Arido_2025} constructed a fine-tuned test in which they were able to show that this fraction can make a difference and lead to a deviation from the true solution (see their Appendix~G). However, for astrophysical set-ups it is not expected to matter. Nevertheless, this effect depends on the size of the time step and becomes weaker for smaller time steps. The same is true for the novel scheme, with the exception that in the traditional schemes the effect is roughly suppressed by the inverse of the neighbour number used to choose the kernel size compared to the scheme presented here. From the test runs presented in Sect.~\ref{sec:tests}, no evidence was found that this effect matters.

\subsection{Scattering-induced subhalo–halo interactions}
\label{sec:methods_SSHI}
For studies focused on substructure evolution, it can be advantageous to resolve only the satellite halos in a simulation while describing the host systems analytically. This allows us to largely reduce the costs compared to a simulation that resolves the host halo as well, especially when focusing on satellites much less massive than the host. In the context of SIDM, this requires not only modelling the gravitational interaction between the host and the satellite halo, but also the self-interactions. To the latter, I refer as scattering-induced subhalo–halo interactions \citep[SSHI,][]{Klemmer_2026}. Developing a scheme for the SSHI was pioneered by \cite{Zeng_2022}. A more general scheme with its implementation in \textsc{OpenGadget3} was introduced by \cite{Klemmer_2026}. Here, I build on this implementation and modify it to allow for the use of the same scattering routine as described above.

The \cite{Klemmer_2026} scheme samples virtual particles, i.e.\ particles that only exist for the purpose of facilitating the SSHI and are discarded afterwards. They are placed at the position of the simulation particle for which the SSHI are computed. These virtual particles represent the local density and local velocity distribution of the host halo.
In contrast to the implementation by \cite{Klemmer_2026}, sampling them at the same position as the $N$-body particle is not an option, as the positions are crucial for the scattering rate. Instead, I sample them within a sphere around the simulation particle. In total,
\begin{equation}
    N_\mathrm{sample} = \left\lceil \frac{4\uppi \, \rho(\mathbf{x}_i) \, r^3_\mathrm{search}}{3 \, m_{\mathrm{n},i}} \right\rceil
\end{equation}
virtual particles are sampled for every simulation particle. Here, $r_\mathrm{search}$ is given by Eq.~\eqref{eq:search_radius}.
Those particles are placed within a radius of
\begin{equation}
    r_\mathrm{sample} = \sqrt[3]{\frac{3 \, N_\mathrm{sample} \, m_{\mathrm{n},i}}{4 \uppi \, \rho(\mathbf{x}_i)}}
\end{equation}
around the corresponding $N$-body particle. After one of the particles is sampled, the scattering computations as described in Sect.~\ref{sec:methods_largeangle}--\ref{sec:methods_successive_interactions} are executed. Afterwards, the next virtual particle is sampled. This continues until all $N_\mathrm{sample}$ virtual particles have been sampled for the considered simulation particle. Subsequently, the code continues with the next simulation particle. In practice, I allow the simulation particles to be treated in parallel as their computations are fully independent.

The implementation in \textsc{OpenGadget3} enables us to model the SSHI for various SIDM models, including full differential cross-sections. In Sect.~\ref{sec:tests_SSHI}, I consider the case of isotropic velocity-independent scattering and demonstrate that the modified implementation gives the expected results for a test set-up.

\section{Verification tests} \label{sec:tests}

In this section, I present four test problems to verify that the implementation works as expected.
Many of the test problems presented in earlier work \citep{Fischer_2021a, Fischer_2022, Fischer_2026b, Patil_2026} have been used to test various aspects of the implementation. They were all passed successfully and a few of them are presented in the following.

\subsection{Scatter count and velocities}

\begin{figure*}
    \centering
    \includegraphics[width=\linewidth]{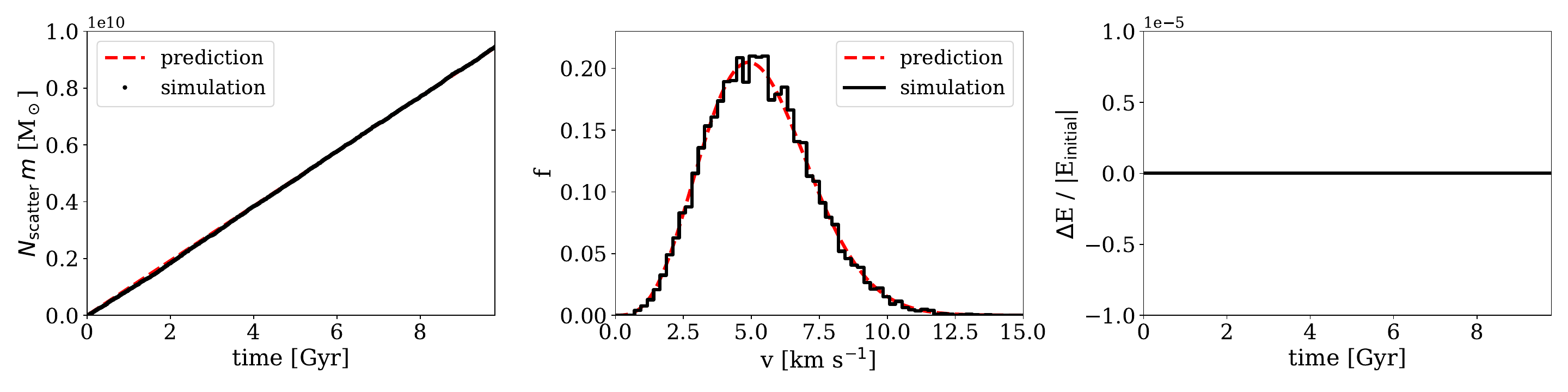}
    \caption{Test problem for the scattering rate and velocities of isotropic, velocity-independent scattering. The left panel gives the number of scatter events times the physical DM particle mass since the start of the simulation (black) and compares it to the analytic solution (red, Eq.~\eqref{eq:scatter_count}). The middle panel displays the distribution of relative velocities at which the scatterings take place (black). Here, the analytic expectation (red) is given by Eq.~\eqref{eq:scatter_vel_dist}. The energy conservation is shown in the right panel. This test was originally introduced by \cite{Fischer_2026b}.}
    \label{fig:test_scatter_count}
\end{figure*}

The first test that I present checks the scattering rate, the scattering velocities, and the energy conservation for isotropic velocity-independent scattering. 
I use a periodic box with a side length of $10\,\mathrm{kpc}$ filled with DM following a constant density ($\rho = 10^7 \, \mathrm{M_\odot} \, \mathrm{kpc}^{-3}$) and a velocity distribution described by a Maxwell--Boltzmann distribution. The one-dimensional velocity dispersion is $\nu = 2\,\mathrm{km}\,\mathrm{s}^{-1}$. The initial conditions (ICs) are sampled with $N = 10^4$ particles. The system is evolved without gravity but under the influence of self-interactions with a cross-section of $\sigma/m = 20 \, \mathrm{cm}^2 \, \mathrm{g}^{-1}$. Given that the simulation particles initially follow a Maxwell--Boltzmann distribution, the system does not evolve due to the scattering but resides in an equilibrium state. This test was first presented in \cite{Fischer_2026b}.

For this test, the expected number of scattering events of physical particles as a function of time is 
\begin{equation} \label{eq:scatter_count}
    N_\mathrm{scatter}(t) = \sqrt{\frac{4}{\uppi}} \frac{\sigma}{m} \, \rho \, \nu \, \frac{N \, m_\mathrm{n}}{m} \, t \,.
\end{equation}
The numerical particle mass is given by $m_\mathrm{n}$. 
One can also compute at which relative velocities, $v_\mathrm{scat}$, these scatterings are expected to occur. The corresponding distribution for a velocity-independent cross-section is 
\begin{equation} \label{eq:scatter_vel_dist}
    f(v_\mathrm{scat}) = \frac{1}{8} \, \frac{v_\mathrm{scat}^3}{\nu^4} \, \exp\left(-\frac{v_\mathrm{scat}^2}{4 \, \nu^2}\right) \,.
\end{equation}
Lastly, I check the conservation of energy. Given that the test employs elastic scattering, the sum of the kinetic energy of the particles should be conserved. 

In Fig.~\ref{fig:test_scatter_count}, I present the results of the test. The left panel shows the scattered mass and compares the simulation results with the expectation from Eq.~\eqref{eq:scatter_count}. The middle panel displays the distribution of velocities at which the interactions took place over the course of the simulation and shows it together with the solution from Eq.~\eqref{eq:scatter_vel_dist}. Lastly, the right panel gives the error in energy conservation relative to the total initial energy. In all three panels, it is clearly visible that the simulation results agree well with the analytic expectations.

Finally, I note that the default formulation without enlarging the interaction radius (Sect.~\ref{sec:methods_successive_interactions}) was sufficient to obtain accurate results. Hence, the interaction radius as given by Eq.~\eqref{eq:interaction_radius_rare} was used only.

\subsection{Deceleration test}

\begin{figure}
    \centering
    \includegraphics[width=\columnwidth]{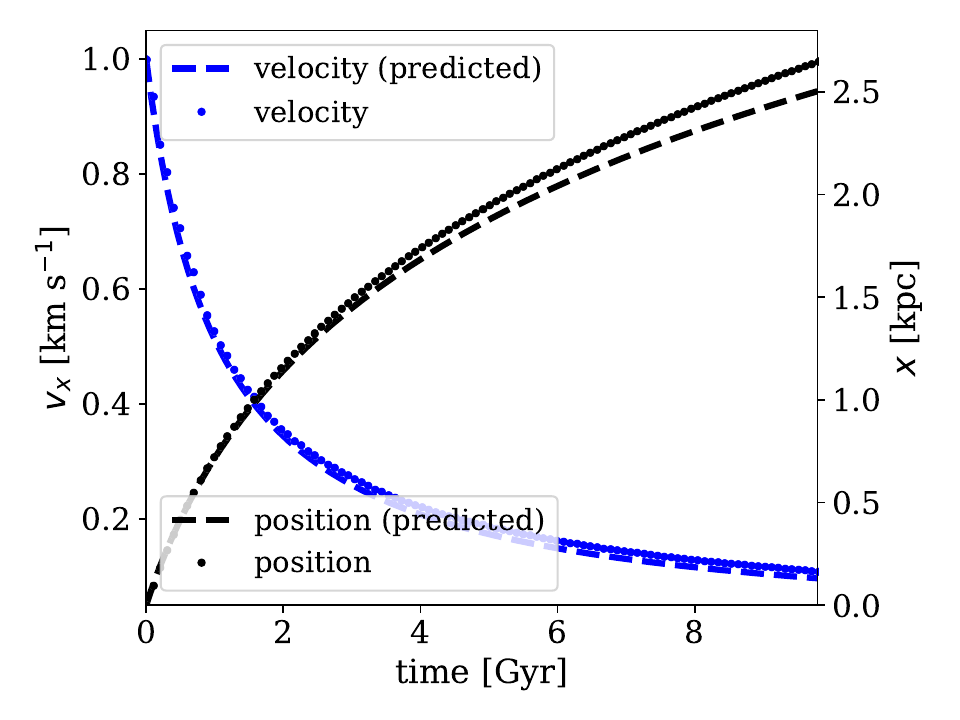}
    \caption{Deceleration test for the small-angle scattering scheme. A single particle travels through a constant background density. Only the drag force is applied while the random component is neglected. The velocity and position of the test particle are shown as a function of time for the simulation and analytic prediction. This test was originally introduced by \cite{Fischer_2021a}.}
    \label{fig:deceleration_problem}
\end{figure}

Besides testing large-angle scattering, I also tested the scheme for small-angle scattering described in Sect.~\ref{sec:methods_smallangle}. Here, I use a test set-up, where a single test particle travels through a constant background density. The velocities of the background particles are zero. In detail, this is the same test as described in sect.~3.1.1 by \cite{Fischer_2021a}. The ICs contain $10^4$ particles, giving rise to a background density of $4.46 \times 10^7 \, \mathrm{M_\odot} \, \mathrm{kpc}^{-3}$, and the test particle has an initial velocity of $1 \, \mathrm{km} \, \mathrm{s}^{-1}$ moving along the $x$-axis. The set-up is evolved with a cross-section of $\sigma_\mathrm{\Tilde{T}} / m = 200 \, \mathrm{cm}^2 \, \mathrm{g}^{-1}$. Importantly, only the drag force is applied, and the random component is neglected. As a consequence, the direction of the particle does not change, only the magnitude of its velocity. In detail, I used, $\delta = 2 \times 10^{-4}$. At the same time, there was no need to enlarge the interaction distance.

The analytic solution, describing the trajectory of the test particle, is given by
\begin{equation}
    \Ddot{x} = -\frac{1}{2} \, \Dot{x}^2 \, \rho \,\frac{\sigma_\mathrm{\Tilde{T}}}{m} \,.
\end{equation}
Here, $x$ denotes the $x$-component of the particle's position.

The simulation results are shown together with the analytic solution in Fig.~\ref{fig:deceleration_problem}.
One can see that the simulation results agree reasonably well with the predicted evolution. The velocity (blue) matches the expectation better than the particle's position (black). These results are very similar to what was found by \cite{Fischer_2021a}.

\subsection{Comoving integration test}

\begin{figure}
    \centering
    \includegraphics[width=\columnwidth]{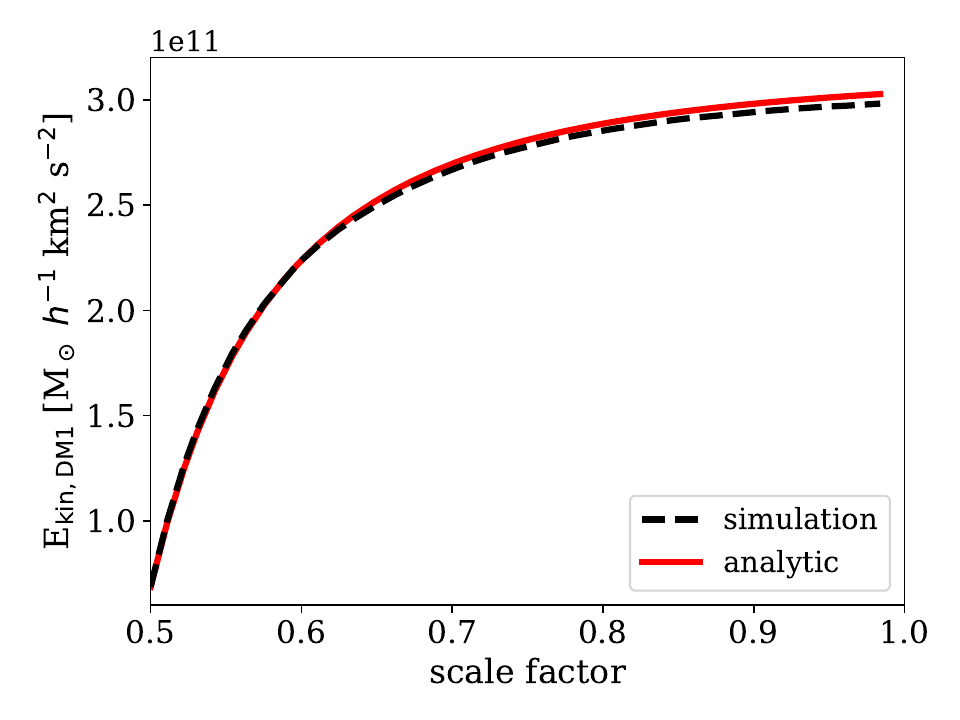}
    \caption{Comoving integration test for isotropic scattering. Heat flows from a hot DM component to a colder one while the space is expanding. The kinetic energy of the colder component is shown as a function of the scale factor.
    Here, the energy is given in terms of the canonical momentum, such that it would be constant if no DM scatterings occur.
    The simulation results are given in black and compared to the analytic expectation (red). This test was originally introduced by \cite{Fischer_2026b}.}
    \label{fig:comoving_integration_test}
\end{figure}

In the following, I test the comoving integration for isotropic scattering. The test problem consists of a periodic box that contains a cold and hot DM component.
Their densities are constant, and their velocity distributions follow a Maxwell--Boltzmann distribution.
The initial peculiar energy of the cold component is $E_\mathrm{DM1} = 2.75 \times 10^{11} \, \mathrm{M_\odot} \,  h^{-1} \, \mathrm{km}^2 \, \mathrm{s}^{-2}$ and the hot component has a peculiar energy of $E_\mathrm{DM2} = 2.75 \times 10^{12} \, \mathrm{M_\odot} \,  h^{-1} \, \mathrm{km}^2 \, \mathrm{s}^{-2}$. Each DM component is sampled with $10^5$ particles. The simulation starts at a scale factor of $a = 0.5$ and is evolved until $a = 1.0$. This is the same test set-up as presented in sect.~3.1.4 by \cite{Fischer_2026b} and was adopted from the DM-baryon study by \cite{Fischer_2025a}.

However, in contrast to these studies, I consider isotropic scattering with a momentum transfer cross-section of $\sigma_\mathrm{T}/m = 2 \times 10^6 \, \mathrm{cm}^2 \, \mathrm{g}^{-1}$. Here, the momentum transfer cross-section is
\begin{equation} \label{eq:momentum_transfer_cross_section}
    \sigma_\mathrm{T} = \int_{-1}^1 \frac{\mathrm{d}\sigma}{\mathrm{d}\cos\theta_\mathrm{cms}} \, (1 - \cos\theta_\mathrm{cms}) \, \mathrm{d}\cos\theta_\mathrm{cms} \,.
\end{equation}
This is the quantity relevant for the heat exchange between the two components and is more meaningful in this case compared to $\sigma_\mathrm{\Tilde{T}}$, as the particles are distinguishable. In addition, enlarging the interaction radius as described in Sect.~\ref{sec:methods_successive_interactions} is necessary for this test.

For deriving the expected solution, I assume that the DM particles of each component follow a Maxwell-Boltzmann distribution. This is true for the ICs but may change over the course of the simulation, as the interactions between the two components introduce corresponding deviations. However, scatterings within each component are also modelled and bring each component closer to a Maxwell-Boltzmann distribution. Overall, assuming a Maxwell-Boltzmann distribution appears to be unproblematic.

The analytic solution follows from \cite{Dvorkin_2014}. In the non-expanding case, the power $P_\mathrm{DM1}$ describing how the energy of one of the two DM components changes over time is given by\footnote{Here, I assume that all DM particles have the same mass. For a more general version, see section~3.1.4 by \cite{Fischer_2026b}.}
\begin{align} \label{eq:heat_conduction_static}
    P_\mathrm{DM1} =& \frac{\mathrm{d}E_\mathrm{DM1}}{\mathrm{d}t} \nonumber\\
    =& -\frac{32}{\sqrt{27\uppi}} \, \frac{\rho_\mathrm{DM1} \, \rho_\mathrm{DM2}}{4} \frac{\sigma_\mathrm{T}}{m} \left(\frac{E_\mathrm{DM1}}{M_\mathrm{DM1}} + \frac{E_\mathrm{DM2}}{M_\mathrm{DM2}}\right)^{1/2} \nonumber \\
    &\times \left[ E_\mathrm{DM1} \left( \frac{1}{\rho_\mathrm{DM1}} + \frac{1}{\rho_\mathrm{DM2}} \right) - \frac{E_\mathrm{tot}}{\rho_\mathrm{DM2}} \right] \,.
\end{align}
Here, $M_\mathrm{DM1}$ and $M_\mathrm{DM2}$ refer to the total mass of each component.
In the case of an expanding space, the differential equation describing the evolution of the system becomes
\begin{equation} \label{eq:heat_conduction_cosmo}
    \frac{\mathrm{d}E_\mathrm{DM1}}{\mathrm{d}a} = \frac{P_\mathrm{DM1}(a)}{a \, H(a)} -\frac{2 E_\mathrm{DM1}}{a} \,.
\end{equation}
Here, $H(a)$ is the Hubble parameter and $P_\mathrm{DM1}(a)$ is given by Eq.~\eqref{eq:heat_conduction_static}. 
The quantities in Eq.~\eqref{eq:heat_conduction_static} and~\eqref{eq:heat_conduction_cosmo} are in terms of their peculiar values, i.e.\ $\rho \propto a^{-3}$ and $E \propto a^{-2}$.
To obtain the solution, I integrate Eq.~\eqref{eq:heat_conduction_cosmo} numerically.

The simulation results are shown in Fig.~\ref{fig:comoving_integration_test} and compared to the analytic solution described above. It is visible that the total kinetic energy of the colder DM component in the simulation agrees well with its expected value.

\subsection{Scattering-induced subhalo–halo interaction test} \label{sec:tests_SSHI}

\begin{figure}
    \centering
    \includegraphics[width=\columnwidth]{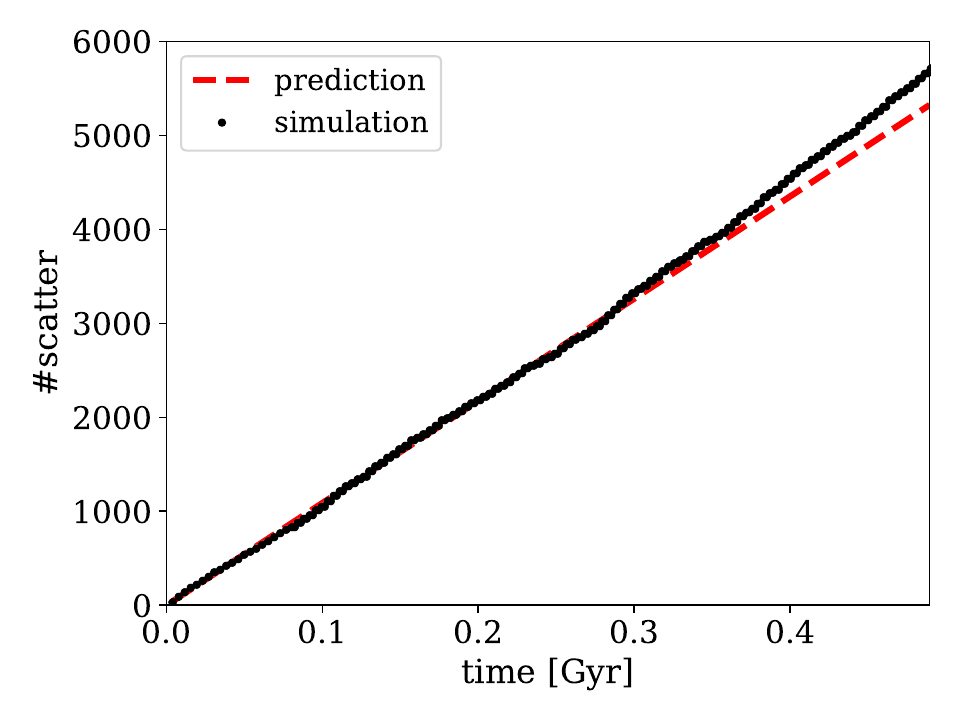}
    \caption{Test for the SSHI implementation. The number of interactions of the satellite particles (black) is shown as a function of time. As a comparison, I show the expected number of interactions (red) according to Eq.~\eqref{eq:SSHI_rate} as well.}
    \label{fig:test_SSHI}
\end{figure}

Here, I test the SSHI implementation described in Sect.~\ref{sec:methods_SSHI}. For this, I simulate a satellite halo within a host system and compare the scattering rate to its expected value.

The analytically described host system consists of DM only and is described by an Navarro-Frenk-White \citep[NFW,][]{Navarro_1996} profile,
\begin{equation} \label{eq:nfw_profile}
    \rho_\mathrm{NFW}(r) = \frac{\rho_0}{\frac{r}{r_\mathrm{s}}\left(1+\frac{r}{r_\mathrm{s}}\right)^2} \, .
\end{equation}
Here, $\rho_0$ denotes the density parameter of the halo and $r_\mathrm{s}$ is its scale radius. For the host system, I choose $\rho_{0, \, \mathrm{host}} = 4.4 \times 10^6 \, \mathrm{M}_\odot \, \mathrm{kpc}^{-3}$ and $r_{\mathrm{s}, \, \mathrm{host}} = 51 \, \mathrm{kpc}$. This corresponds to a halo with a virial mass of $M_\mathrm{vir} \approx 10^{13} \, \mathrm{M_\odot}$ and a typical concentration of $c_\mathrm{vir} \approx 8.8$ of the mass-concentration relation \citep{Dutton_2014}.

The satellite halo is chosen to be much less massive. This has the advantage that dynamical friction can be neglected, but important for the test is that the internal velocity dispersion of the system hardly affects the scattering rate.
The NFW halo parameters are $\rho_{0, \, \mathrm{sat}} = 1.7 \times 10^6 \, \mathrm{M_\odot} \, \mathrm{kpc}^{-3}$ and $r_{\mathrm{s}, \, \mathrm{sat}} = 0.2 \, \mathrm{kpc}$. This gives a halo mass of $\approx 2 \times 10^5 \, \mathrm{M_\odot}$.
I use \textsc{SpherIC} \citep{Garrison-Kimmel_2013} to generate the ICs using $N=10^5$ particles.
The subhalo is placed at $r_\mathrm{sat} = 2.163 \, r_{\mathrm{s},\,\mathrm{host}}$ on a circular orbit with a velocity of $v_\mathrm{sat} = 1.649  \, r_{\mathrm{s},\, \mathrm{host}} \, \sqrt{\mathrm{G} \, \rho_{0,\, \mathrm{host}}}$. This choice corresponds to the maximum of the rotation curve.

I estimate the rate at which a satellite particle scatters with a host particle to be
\begin{equation} \label{eq:SSHI_rate}
    R = \rho_\mathrm{host} \, \nu_\mathrm{host} \, \frac{\sigma}{m} \, \sqrt{2} \, \left[\left(x + \frac{1}{2 \, x} \right) \, \mathrm{erf}(x) + \frac{e^{-x^2}}{\sqrt{\uppi}} \right] \, .
\end{equation}
Here, I use $x = v_\mathrm{sat} / \left(\sqrt{2} \, \nu_\mathrm{host}\right)$, and $\rho_\mathrm{host}$ denotes the local density of the host halo, whereas $\nu_\mathrm{host}$ is its local one-dimensional velocity dispersion. 
For the sake of simplicity, I assume that the velocities of the halo locally follow a Maxwell--Boltzmann distribution. Moreover, I neglect the velocity dispersion of the subhalo.

I simulate the halo for almost $0.5 \, \mathrm{Gyr}$ assuming isotropic and velocity-independent scattering with a total cross-section per DM mass of $\sigma/m = 10 \, \mathrm{cm}^2 \, \mathrm{g}^{-1}$. 
For this simulation, it was not necessary to enlarge the interaction distance as described in Sect.~\ref{sec:methods_successive_interactions}.
Throughout the simulation, I keep track of the total number of scatterings that occur. Given the small velocity dispersion of the satellite halo, interactions between particles belonging to the satellite are very rare compared to the SSHI.

In Fig.~\ref{fig:test_SSHI}, I show the results for my test set-up. It is visible that the simulation agrees reasonably well with the scattering rate according to Eq.~\eqref{eq:SSHI_rate}. Given the approximations made, small deviations are expected, which become mainly visible at the later times shown.

\section{Strong scaling tests of the implementation} \label{sec:scaling}

In this section, I show strong scaling tests for the early evolution of an isolated DM halo (Sect.~\ref{sec:scaling_halo}). 
Here, I test the parallelisation for shared and distributed memory architectures based on Open Multiprocessing (OpenMP) and the  Message Passing Interface (MPI).
The parallelisation strategy for the SIDM computations has been described in detail in sect.~2.12 by \cite{Fischer_2026b}.
Moreover, I also perform a test for the MPI parallelisation using a cosmological simulation (Sect.~\ref{sec:scaling_cosmo}).

\subsection{Isolated halo} \label{sec:scaling_halo}
\begin{figure}
    \centering
    \includegraphics[width=\columnwidth]{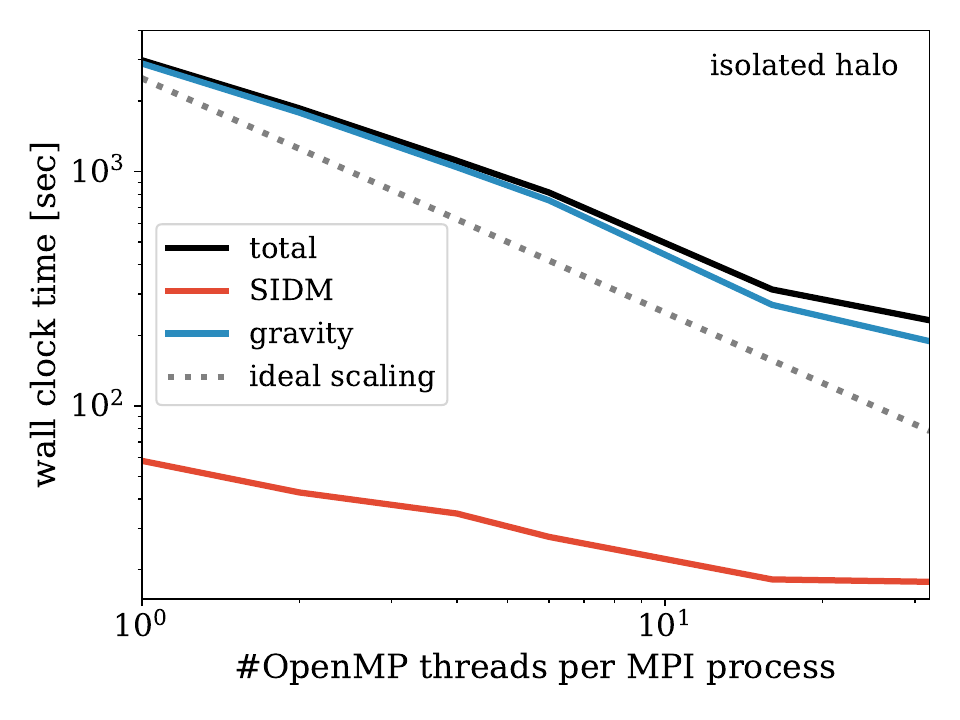}
    \caption{Strong scaling test for the OpenMP parallelisation using an isolated DM halo. The wall clock time after simulating the DM halo for almost 10 Myr is shown as a function of the number of employed OpenMP threads. The number of MPI ranks was fixed to 18, and the total number of particles is $N=10^7$. The total computational costs are shown (black) along with those for the gravity (blue) and self-interaction (red) computations. The gray dotted line indicates a scaling proportional to the inverse of the number of threads.}
    \label{fig:strong_scaling_openmp}
\end{figure}

\begin{figure}
    \centering
    \includegraphics[width=\columnwidth]{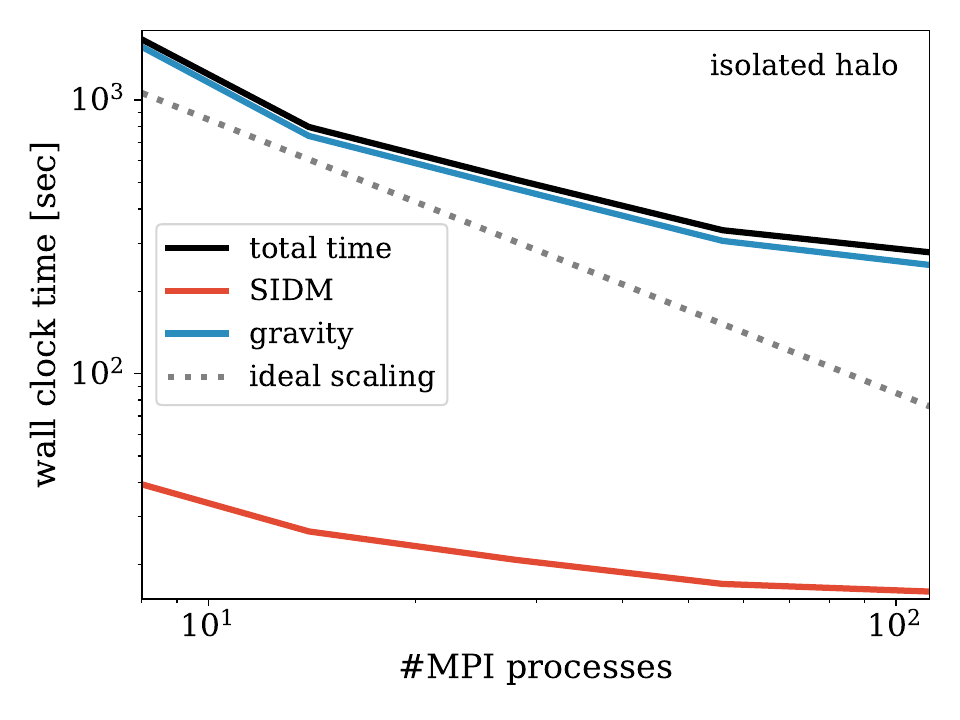}
    \caption{Strong scaling test for the MPI parallelisation using an isolated DM halo. The wall clock time is displayed as a function of the number of employed MPI ranks. The different curves show the total computational costs (black), the ones for SIDM (red), and the ones for gravity (blue). Four OpenMP threads were used for the simulations.}
    \label{fig:strong_scaling_mpi}
\end{figure}

For the first scaling tests, I used an isolated DM halo initially following an NFW profile (Eq.~\eqref{eq:nfw_profile}) with an additional truncation at $15 \, r_\mathrm{s}$. Here, I closely followed the scaling tests in \cite{Fischer_2026b} using the same initial conditions.
The halo parameters are $\rho_0 = 4.42 \times 10^7 \, \mathrm{M_\odot} \, \mathrm{kpc}^{-3}$ and $r_\mathrm{s} = 1.28 \, \mathrm{kpc}$, and the $N = 10^7$ particles were sampled with \textsc{SpherIC} \citep{Garrison-Kimmel_2013}. For the simulations I employed a gravitational softening length of  $\epsilon = 3.0 \, \mathrm{pc}$ and choose a velocity-independent isotropic cross-section of $\sigma/m = 30 \, \mathrm{cm}^2 \, \mathrm{g}^{-1}$. The halo is evolved for almost $10\,\mathrm{Myr}$.

In Fig.~\ref{fig:strong_scaling_openmp}, I show how the simulation code scales when varying the number of OpenMP threads per MPI process while the number of MPI processes is kept fixed at 18.
The total computational costs together with the ones made up by gravity and SIDM are displayed.\footnotemark\setcounter{FootnoteA}{\value{footnote}} \footnotetext[\value{FootnoteA}]{The computational costs have been measured using the SPACE-Timers by \cite{Karademir_2026}.}
It is visible that the computational costs fall behind the ideal scaling of being inversely proportional to the number of OpenMP threads. Nevertheless, increasing the number of threads leads to a substantial speed-up of the simulation, and the code scales well until a relatively high number of OpenMP threads. Especially the scaling for the SIDM costs has substantially improved compared to \cite{Fischer_2026b}, which is likely a consequence of having a much smaller neighbour list that the code needs to iterate over. Hence, one can reasonably make use of many more OpenMP threads compared to that earlier work.
Remarkably are the overall computational costs for SIDM compared to the total costs. They make up only a small fraction of the total cost, rising from slightly below 2\% when one thread is used to a little below 6\% for 16 threads, and for 32 threads the relative fraction still lies below 9\%.

Next, I test the scaling of the MPI parallelisation. The number of OpenMP threads is kept fixed at 4 threads per MPI rank. Again, the wall clock time is measured once the halo has evolved for almost $10 \, \mathrm{Myr}$.\footnotemark[\value{FootnoteA}] The results are shown in Fig.~\ref{fig:strong_scaling_mpi}. Similar to the test for the OpenMP parallelisation, the scaling falls behind the ideal behaviour but nevertheless scales relatively well, making the implementation well suited for massively parallelised computations. As before, the very low fraction of the SIDM computational cost demonstrates the performance advantage of the novel scheme.

\subsection{Cosmological box} \label{sec:scaling_cosmo}

\begin{figure}
    \centering
    \includegraphics[width=\columnwidth]{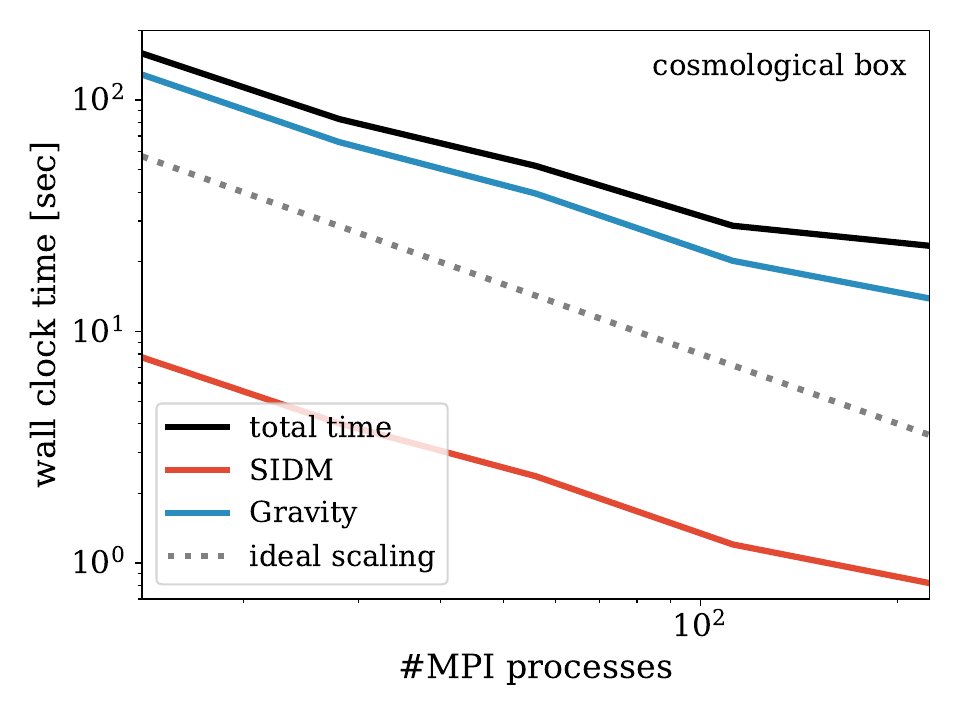}
    \caption{Strong scaling test for the MPI parallelisation using a cosmological box. The wall clock time is shown as a function of the number of MPI processes. The total computational costs (black), the ones for SIDM (red), and the ones for gravity (blue) are indicated separately. The simulations were run employing four OpenMP threads.}
    \label{fig:scaling_cosmo}
\end{figure}

In addition to the test for an isolated DM halo, I also use a cosmological set-up, namely the same as employed by \cite{Fischer_2026b}. It is a DM-only simulation with $N=576^3$ particles of equal mass in a cubic box with a comoving side length of $l_\mathrm{box} = 48 \, \mathrm{cMpc} / h$.
The following cosmological parameters are employed: $\Omega_\mathrm{M} = 0.272$, $\Omega_\Lambda = 0.728$, $h = 0.704$, $n_\mathrm{s} = 0.963$, and $\sigma_8 = 0.809$~\citep[WMAP7,][]{Komatsu_2011}.
To generate the ICs, \textsc{N-GenIC}~\citep{Springel_2015} was used.
For the SIDM model, I employ isotropic and velocity-independent scattering with a total cross-section of $\sigma/m = 1 \, \mathrm{cm}^2 \, \mathrm{g}^{-1}$.

In Fig.~\ref{fig:scaling_cosmo}, I show the results, namely the wall clock time it takes to evolve the system starting at $z=60$ for a few time steps as a function of the number of MPI processes. As before, one can see that overall the computational costs scale well with the number of MPI ranks.\footnotemark[\value{FootnoteA}] Remarkably, the SIDM costs make up only a small fraction of the total costs, as found for the isolated halo as well. The simulation is clearly limited by the computations required for gravity and not by those for the DM scattering.

\section{Gravothermal halo evolution} \label{sec:halo}

SIDM halos can undergo a process known as gravothermal collapse~\citep[e.g.][]{Balberg_2002}. The self-interactions conduct heat outwards, which makes the central region denser and increases its velocity dispersion due to the negative heat capacity. The increased velocity dispersion enhances the heat conduction and accelerates the evolution. This process is well known from globular clusters~\citep[e.g.][]{Lynden-Bell_1980}.

Gravothermally collapsing halos have gained substantial interest in recent years, motivated by observations that point to fairly compact dark objects~\citep[e.g.][]{Yu_2026}. For example, this concerns substructure in gravitational strong lenses~\citep[e.g.][]{Minor_2020, Enzi_2024, Kollmann_2025, Tajalli_2025, Vegetti_2026} or affects flux ratio anomalies of lensed quasars~\citep[e.g.][]{Gilman_2026a}.

Modelling gravothermally collapsing SIDM halos has been challenging, in particular in $N$-body simulations due to numerical challenges~\citep{Zhong_2023, Mace_2024, Palubski_2024, Fischer_2024b}, and several studies have been conducted to compare different methods~\citep[e.g.][]{Mace_2025, Li_2026, Bosch_2026b, Gurian_2026}.
Despite pioneering work by \cite{Zeng_2026}, a framework that allows simulating SIDM halos with high physical fidelity into the post-collapse phase remains out of reach.

In the following, I investigate how much the novel scheme can contribute to these challenges.
Therefore, I study the gravothermal evolution of an isolated DM halo. I describe the simulation set-up (Sect.~\ref{sec:halo_sim_setup}) followed by the results showing the halo evolving into the core collapse phase (Sect.~\ref{sec:halo_results}).

\subsection{Simulation set-up} \label{sec:halo_sim_setup}
 
I use the same halo profile as before in Sect.~\ref{sec:scaling_halo}.
It corresponds to a DM halo that has a mass of about $2.8 \times 10^9 \, \mathrm{M_\odot}$ and a concentration parameter of $c \approx 22$.
However, here I use a different resolution: the ICs are sampled with $N = 2 \times 10^6$, and a gravitational softening length of $\epsilon = 6.4 \, \mathrm{pc}$ is employed.
For the self-interactions, velocity-independent isotropic scattering is chosen with a total cross-section of $\sigma/m = 80 \, \mathrm{cm}^2 \, \mathrm{g}^{-1}$. 
These choices allow comparison with the results of the previous simulations of \cite{Fischer_2025b}.

In addition to this simulation, I ran lower-resolution simulations. For consistency, the ICs were generated with \textsc{SpherIC} \citep{Garrison-Kimmel_2013}. Here, I created multiple realisations and selected those with a small realisation noise, which can lead to substantial inaccuracies for poorly resolved halos, as pointed out by \cite{Mace_2024} \citep[see also][]{Schmidt_2026}.
The gravitational softening lengths for all simulations of this section are chosen close to
\begin{equation} \label{eq:gravitational_softening}
		\epsilon = \frac{2.3 \, r_\mathrm{s}}{\sqrt{N(r < r_\mathrm{s})}} \,.
\end{equation}
With this choice, I follow Eq.~(C.3) by \cite{Schmidt_2026}.
In detail, the selected particle numbers and softening lengths are $N \in \{3.2 \times 10^3$, $1.6 \times 10^4$, $8 \times 10^4$, $4 \times 10^5\}$ and $\epsilon \in 
\{160~\mathrm{pc}$, $72~\mathrm{pc}$, $32~\mathrm{pc}$, $14~\mathrm{pc}\}$ respectively.

The opening criterion for the tree when computing the gravitational forces used in \textsc{OpenGadget3} has been described by \citet[][as expressed in his Eq.~(18)]{Springel_2005}. For the simulation, I set the corresponding parameter to \texttt{ErrTolForceAcc} = $5 \times 10^{-4}$.

Lastly, the simulations here are run without using an extended interaction distance described in Sect.~\ref{sec:methods_successive_interactions}. Especially in the early evolution stage, there is no need to do so. Only at late stages do successive interactions occur more frequently. The decision not to increase the interaction distance also helps to keep the overall computational costs low; the costs for the SIDM computations over the full simulation do not exceed $5\%$ of the total simulation costs.\footnotemark[\value{FootnoteA}]

\subsection{Results} \label{sec:halo_results}

\begin{figure}
    \centering
    \includegraphics[width=\columnwidth]{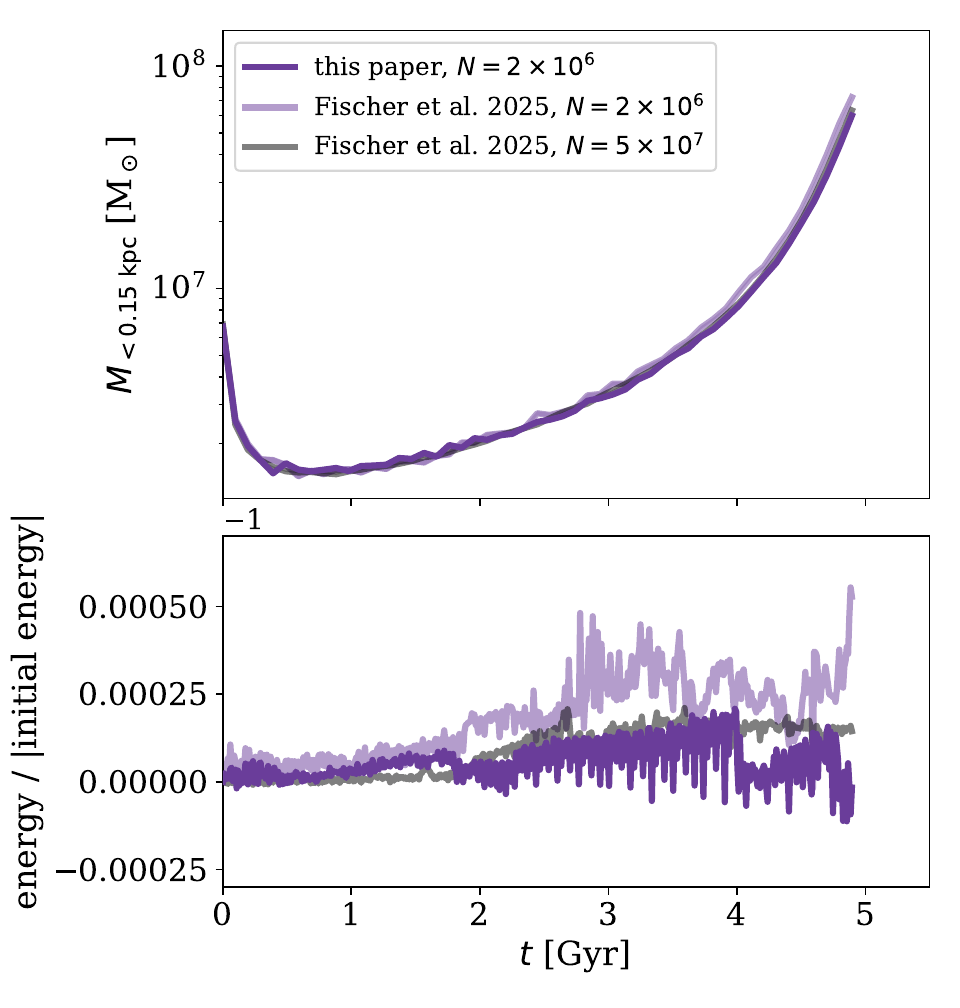}
    \caption{Gravothermal evolution of an isolated DM halo. The upper panel gives the mass enclosed within $0.15 \, \mathrm{kpc}$ as a function of time. The lower panel gives the total energy relative to the absolute of its initial value. The solid line shows the results with the novel scheme and the faint ones give the results of simulation J (purple) and Q (black) by \cite{Fischer_2025b}.}
    \label{fig:halo_central_density}
\end{figure}

\begin{figure}
		\centering
		\includegraphics[width=\columnwidth]{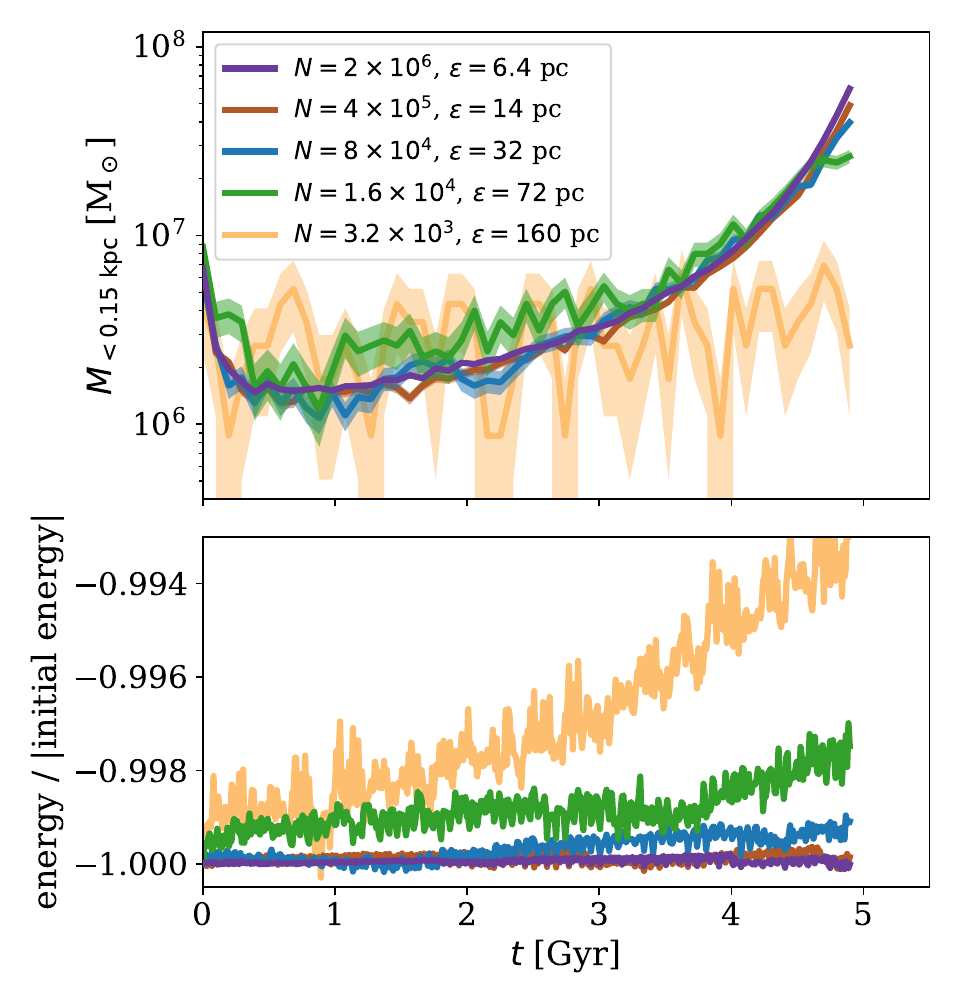}
		\caption{Gravothermal evolution of an isolated DM halo for low resolutions. The same as in Fig.~\ref{fig:halo_central_density} is shown, but now the particle number and softening length are varied. All simulations were run with the new scheme. The purple simulation is the same as shown in Fig.~\ref{fig:halo_central_density}.}
		\label{fig:halo_central_density_low_res}
\end{figure}

In the following, I compare the simulation results obtained with the novel scheme to previous simulations by \cite{Fischer_2025b}. I identify the centre of the halo using the peak-finding algorithm by \cite{Fischer_2021b} and compute the mass around it enclosed in a sphere with a radius of $r = 0.15 \, \mathrm{kpc}$. 

The enclosed mass as a function of time is shown in the upper panel of Fig.~\ref{fig:halo_central_density}, while the lower panel gives the total energy (kinetic + potential) relative to the absolute value of the initial total energy as a function of time. The result by \cite{Fischer_2025b} using the same simulation set-up is shown by the faint purple line. It is visible that the collapse with the novel scheme (solid purple line) evolves slightly slower, but overall agrees fairly well. The relatively slower evolution could be caused by multiple effects. One of them is that a higher spatial resolution that properly resolves the mean free path avoids leading to an artificially enhanced heat conductivity and thus would result in a more accurate evolution. Another aspect is the successive scatterings described in Sect.~\ref{sec:methods_successive_interactions}; they would artificially slow down the evolution. Moreover, one can see that the novel scheme leads to better energy conservation, which would actually speed up the evolution compared to the older simulation by \cite{Fischer_2025b}. The difference in the energy conservation likely stems from differences in the time steps. In the new simulation, the time steps are set by the gravitational time step criterion only. This avoids populating the smallest time bins used in the older simulation, and likely time step changes are less frequent. The latter helps to improve energy conservation, as the time step criteria are not time-symmetrically evaluated and thus give rise to energy errors \citep{Dehnen_2017, Fischer_2024b}.

In Fig.~\ref{fig:halo_central_density}, I also show the highest-resolved and plausibly most accurate simulation by \cite{Fischer_2025b}. It uses a smaller gravitational softening length of $\epsilon = 1 \, \mathrm{pc}$. In principle, the gravitational softening leads to an artificially shallower potential compared to the true physical solution, leading to a too-slow gravothermal evolution. However, this effect is tiny and not of concern here, as the new simulation and the highest resolved one by \cite{Fischer_2025b} agree fairly well. Interestingly, the new simulation gives comparable or even slightly better energy conservation than the older one with a 25 times higher resolution. This is likely related to the time steps as explained above.

In terms of the accuracy of the results, there is no clear advantage of the $5\times10^7$ particle simulation of \cite{Fischer_2025b} over the new $2 \times 10^6$ particle simulation for the simulated time span, except that estimates of quantities such as the density are less noisy.
The collapse time between the two simulations deviates by only $0.41 \%$.\footnotemark\setcounter{FootnoteB}{\value{footnote}} \footnotetext[\value{FootnoteB}]{Here, I compared the times it takes to reach an enclosed mass of $M(r<0.15 \, \mathrm{kpc}) = 1.4 \times 10^7 \, \mathrm{M_\odot}$.}

Additionally, I ran tests varying the number of simulation particles, choosing the gravitational softening length close to Eq.~\eqref{eq:gravitational_softening}.
The results are shown in Fig.~\ref{fig:halo_central_density_low_res}.
For the higher resolution runs shown ($N \geq 8 \times 10^4$), the gravothermal evolution agrees fairly well among the simulations. Even the $N = 1.6 \times 10^4$, which is much noisier due to the low particle number, provides a fairly accurate collapse time. However, this is to some extent by chance.
Only the simulation with $3.2 \times 10^3$ particles fails to properly capture the gravothermal evolution, and given its poor resolution, the results are very noisy.

Moreover, it is important that the intrinsic randomness of the self-interactions also has an impact on the collapse times. Reruns of a simulation do not result in exactly the same collapse time, where the spread of the resulting collapse time distribution depends on the resolution. For the highly resolved simulations, it is fairly small and not a concern.
However, further tests for the $N = 1.6 \times 10^4$ case revealed a deviation of up to $\approx 12\%$ from the collapse time for 10 realisations of the simulation.\footnotemark[\value{FootnoteB}]
Here, simulation Q by \cite{Fischer_2025b} was used as a reference for the collapse time. 
In the $N = 1.6 \times 10^4$ simulations, the particle mass is about $1.74 \times 10^5 \, \mathrm{M_\odot}$.

In conclusion, the new scheme allows one to achieve comparable accuracy to the \cite{Fischer_2025b} results at a much lower resolution. Moreover, given a specific resolution, it allows one to achieve more accurate results at reduced computational costs. Finally, the new scheme gives relatively accurate results for low-resolved halos ($N\gtrsim 8 \times 10^4$), and thus enhances our ability to simulate the gravothermal collapse of SIDM halos. This is particularly relevant for cosmological simulations as halos are typically poorer resolved compared to idealised studies.

\section{Discussion} \label{sec:discussion}

In this section, I discuss various aspects of the present numerical scheme to simulate DM self-interactions. This concerns, in addition to other aspects, its limitations and open questions.

\subsection{Is this a Monte Carlo scheme?}

Although the scheme presented here does not explicitly draw random or pseudo-random numbers to decide if particles interact, it is nevertheless crucially based on randomness.
The particle distribution is regarded as randomly sampled from the true underlying distribution. Based on this, the interaction criteria for the numerical particles have been derived (see Eqs.~\eqref{eq:interaction_radius_rare} and~\eqref{eq:interaction_radius_freq}). The required randomness in the particle positions enters initially when the ICs for the simulation are generated and serves the numerical scheme later on.

Nevertheless, pseudo-random numbers are still used in a Monte Carlo fashion during the simulation. The scattering angles are not determined by the particle positions, but sampled based on the differential cross-section using pseudo-random numbers.

\subsection{Limitations of spatial resolution}

In principle, the scheme presented here allows achieving an arbitrarily high spatial resolution for the particle interactions when decreasing the time step size accordingly.
However, as discussed in Sect.~\ref{sec:methods_successive_interactions}, this leads to the problem of successive interactions. When the same pair interacts multiple times without any of the other particles having another interaction in between, this artificially suppresses the effect of the self-interactions.

Avoiding problems related to successive interactions effectively limits the spatial resolution. In traditional SIDM schemes where the kernel size is chosen adaptively, the spatial resolution is limited by the mean particle separation or a small multiple of it set by the number of the chosen neighbours. In contrast, the new scheme can achieve a higher spatial resolution in typical astrophysical set-ups, when Eq.~\eqref{eq:enlarge_crit} is not fulfilled. This is, for example, the case for a small cross-section.

\subsection{Comparison to previous schemes for SIDM}

Modern SIDM implementations typically follow a scheme similar to the one introduced by \cite{Kochanek_2000}. Here, interaction partners are chosen within the vicinity of the particle within a length set by a nearest-neighbour number. Then interacting particles are chosen probabilistically by drawing pseudo-random numbers. The schemes typically used differ in how interaction probabilities are computed and used to determine which particles interact with each other. As far as I am aware, the only schemes that do not use a probabilistic approach are the ones by \cite{Huo_2019}, the one for small-angle scattering by \cite{Fischer_2021a}, and the one introduced here.

Modern SIDM schemes usually search for the nearest $N_\mathrm{ngb}$ neighbouring particles to enable the distance within which particles interact to be adaptive. Finding these neighbouring particles comes with substantial computational costs. These computations have become superfluous in the novel scheme described here. This, together with the fact that very few or even no neighbouring particles need to be considered for the interactions, helps to enhance the performance.
However, for the small-angle scattering scheme, one needs to consider more neighbouring particles than for isotropic scattering. This makes it more expensive than large-angle scattering. Nevertheless, the computational costs remain much smaller than those of gravity.

Notably about the new scheme is that it does not involve any form of a density estimate as far as its core formulation is concerned. However, the mitigation strategy for successive interactions (Sect.~\ref{sec:methods_successive_interactions}) does require the local density. But here it is sufficient to use a rough estimate based on the tree.

Moreover, the new scheme did not change how the scattering kinematics are handled. This makes it compatible with many extensions that have been developed for various DM models.
This applies, for example, to the scheme by \cite{Arido_2025} for combining the methods for small- and large-angle scattering, as well as to SIDM models with dissipation~\cite{Schmidt_2026}.

\subsection{Open questions}
Several open questions and directions for further work remain. Here, I would like to highlight the following:
\begin{enumerate}
    \item Advancing the understanding of the limitations arising from successive scattering in greater detail to find an optimal strategy to enlarge the interaction distance for various differential cross-sections.
    \item Investigating in detail in which regimes it would be advantageous to use a separate SIDM time step criterion, developing and testing a formulation for it.
    \item I have implemented a few criteria to assess the accuracy of the simulation on the fly (see Sect.~\ref{sec:methods_successive_interactions}). However, they may not detect all possible problems. For example, I do not check whether the mean free path is resolved. There are several aspects for which dedicated criteria could be developed that would help to ensure accurate simulation results.
    \item Evaluating whether the new scheme gives biased results for simulation set-ups where the particle positions in the ICs are not fully random. This concerns cosmological simulations, where the assumption of random particle positions might be well satisfied in the central region of halos, but whether it is unproblematic for the full evolution is worth investigating.
\end{enumerate}
Beyond these aspects, there exist several more that would allow us to advance the state of the art of $N$-body SIDM simulations in the future.

\section{Conclusion} \label{sec:conclusion}

With this paper, I have introduced a new scheme to model SIDM within $N$-body simulations.
It is distinct from previous schemes as it does not perform a separate neighbour search to determine a kernel size, nor does it use kernel functions. Instead of computing an interaction probability, the scheme is based on the assumption that the particle positions are random. Particles whose distance is below a threshold interact. Through several test problems, I have shown that it reproduces known solutions. The scheme holds the following advantages over state-of-the-art SIDM schemes.
\begin{itemize}
    \item The scheme has improved spatial resolution as interactions occur only between the closest particles. This is advantageous for resolving the mean free path. Moreover, the spatial resolution can be improved by decreasing the simulation time step, as long as one has not reached the limit imposed by successive interactions. In particular, simulations with poorly resolved halos benefit from these advantages.
    \item The computational costs are reduced as several computations of state-of-the-art schemes are no longer required, and fewer neighbouring particles are considered. Importantly, SIDM simulations, as those presented here with isotropic scattering, spend only a few percent of the total computing time on the self-interactions. Hence, the computational costs for SIDM are not a limiting factor, as long as no extreme situation such as being very deeply in the short mean free path regime occurs.
\end{itemize}
As the novel SIDM scheme allows for more accurate and performant simulations than before, it is expected to be valuable to the SIDM community. Moreover, it does not imply any additional constraints in terms of particle physics models that can be simulated. The new implementation in \textsc{OpenGadget3} will be made publicly available in the near future to allow for new simulations fostering physical insights into the evolution of astrophysical systems under the influence of DM self-interactions.

\begin{acknowledgements}
I am thankful for the inspiring and fruitful discussions with various people interested in the numerical schemes for modelling SIDM. In particular, I want to thank Andrew Robertson for his different perspective on the interpretation of the numerical equations for the state-of-the-art probability-based schemes.
I gratefully acknowledge the support of the Alexander von Humboldt Foundation through a Feodor Lynen Research Fellowship.
Moreover, I acknowledge the Leibniz Supercomputing Centre for funding this project by providing computing time on its Linux-Cluster.\\
\\
Software:
NumPy \citep{NumPy},
Matplotlib \citep{Matplotlib}.
\end{acknowledgements}

\section*{Data availability}
The simulation code \textsc{OpenGadget3} is available in a public LRZ GitLab repository (\url{https://gitlab.lrz.de/AstroCodes/OpenGadget3}). In the near future, it will also include the implementation of the novel scheme presented here. Other data underlying this article will be shared upon reasonable request to the author.

\bibliographystyle{aa}
\bibliography{references.bib}

@ARTICLE{NumPy,
    crossref = {Harris_2020}
}

@ARTICLE{Matplotlib,
    crossref = {Hunter_2007}
}

@ARTICLE{Adhikari_2025,
       author = {{Adhikari}, Susmita and {Banerjee}, Arka and {Boddy}, Kimberly K. and {Cyr-Racine}, Francis-Yan and {Desmond}, Harry and {Dvorkin}, Cora and {Jain}, Bhuvnesh and {Kahlhoefer}, Felix and {Kaplinghat}, Manoj and {Nierenberg}, Anna and {Peter}, Annika H.~G. and {Robertson}, Andrew and {Sakstein}, Jeremy and {Zavala}, Jes{\'u}s},
        title = "{Astrophysical tests of dark matter self-interactions}",
      journal = {Reviews of Modern Physics},
         year = 2025,
        month = oct,
       volume = {97},
       number = {4},
          eid = {045004},
        pages = {045004},
          doi = {10.1103/m2vm-59y3},
archivePrefix = {arXiv},
       eprint = {2207.10638},
 primaryClass = {astro-ph.CO},
       adsurl = {https://ui.adsabs.harvard.edu/abs/2025RvMP...97d5004A}
}

@article{Akhlaghi_2023,
title = {A comprehensive review on micro- and nano-scale gas flow effects: Slip-jump phenomena, Knudsen paradox, thermally-driven flows, and Knudsen pumps},
journal = {Physics Reports},
volume = {997},
pages = {1-60},
year = {2023},
note = {A Comprehensive Review on Micro- and Nano-Scale Gas Flow Effects: Slip-Jump Phenomena, Knudsen Paradox, Thermally-Driven Flows, and Knudsen Pumps},
issn = {0370-1573},
doi = {https://doi.org/10.1016/j.physrep.2022.10.004},
url = {https://www.sciencedirect.com/science/article/pii/S0370157322003787},
author = {Hassan Akhlaghi and Ehsan Roohi and Stefan Stefanov}
}

@ARTICLE{Arido_2025,
       author = {{Arido}, Cenanda and {Fischer}, Moritz S. and {Garny}, Mathias},
        title = "{Simulating realistic self-interacting dark matter models including small and large-angle scattering}",
      journal = {\aap},
         year = 2025,
        month = feb,
       volume = {694},
          eid = {A297},
        pages = {A297},
          doi = {10.1051/0004-6361/202452551},
archivePrefix = {arXiv},
       eprint = {2410.07175},
 primaryClass = {astro-ph.CO},
       adsurl = {https://ui.adsabs.harvard.edu/abs/2025A&A...694A.297A}
}

@ARTICLE{Balberg_2002,
       author = {{Balberg}, Shmuel and {Shapiro}, Stuart L. and {Inagaki}, Shogo},
        title = "{Self-Interacting Dark Matter Halos and the Gravothermal Catastrophe}",
      journal = {\apj},
         year = "2002",
        month = "Apr",
       volume = {568},
       number = {2},
        pages = {475-487},
          doi = {10.1086/339038},
archivePrefix = {arXiv},
       eprint = {astro-ph/0110561},
 primaryClass = {astro-ph},
       adsurl = {https://ui.adsabs.harvard.edu/abs/2002ApJ...568..475B}
}

@article{Banerjee_2020,
	doi = {10.1088/1475-7516/2020/02/024},
	url = {https://doi.org/10.1088\%2F1475-7516\%2F2020\%2F02\%2F024},
	year = 2020,
	month = {feb},
	publisher = {{IOP} Publishing},
	volume = {2020},
	number = {02},
	pages = {024--024},
	author = {Arka Banerjee and Susmita Adhikari and Neal Dalal and Surhud More and Andrey Kravtsov},
	title = {Signatures of self-interacting dark matter on cluster density profile and subhalo distributions},
	journal = {\jcap}
}

@article{Bird_1963,
    author = {Bird, G. A.},
    title = {Approach to Translational Equilibrium in a Rigid Sphere Gas},
    journal = {The Physics of Fluids},
    volume = {6},
    number = {10},
    pages = {1518-1519},
    year = {1963},
    month = {10},
    issn = {0031-9171},
    doi = {10.1063/1.1710976},
    url = {https://doi.org/10.1063/1.1710976},
    eprint = {https://pubs.aip.org/aip/pfl/article-pdf/6/10/1518/12508675/1518_1_online.pdf},
}

@book{Bird_1994,
    author = {Bird, G A},
    title = {Molecular Gas Dynamics And The Direct Simulation Of Gas Flows},
    publisher = {Oxford University Press},
    year = {1994},
    month = {05},
    isbn = {9780198561958},
    doi = {10.1093/oso/9780198561958.001.0001},
    url = {https://doi.org/10.1093/oso/9780198561958.001.0001},
}

@ARTICLE{Bosch_2026b,
       author = {{van den Bosch}, Frank C. and {Tokayer}, Yarone M.},
        title = "{Not so fast: revised calibration of SIDM core collapse}",
      journal = {\mnras},
         year = 2026,
        month = dec,
       volume = {553},
       number = {2},
          eid = {stag2015},
        pages = {stag2015},
          doi = {10.1093/mnras/stag2015},
archivePrefix = {arXiv},
       eprint = {2610.28890},
 primaryClass = {astro-ph.CO},
       adsurl = {https://scixplorer.org/abs/2026MNRAS.553g2015V}
}

@ARTICLE{Burkert_2000,
       author = {{Burkert}, Andreas},
        title = "{The Structure and Evolution of Weakly Self-interacting Cold Dark Matter Halos}",
      journal = {\apjl},
         year = 2000,
        month = may,
       volume = {534},
       number = {2},
        pages = {L143-L146},
          doi = {10.1086/312674},
archivePrefix = {arXiv},
       eprint = {astro-ph/0002409},
 primaryClass = {astro-ph},
       adsurl = {https://ui.adsabs.harvard.edu/abs/2000ApJ...534L.143B}
}

@ARTICLE{Carlson_1992,
       author = {{Carlson}, Eric D. and {Machacek}, Marie E. and {Hall}, Lawrence J.},
        title = "{Self-interacting Dark Matter}",
      journal = {\apj},
         year = 1992,
        month = oct,
       volume = {398},
        pages = {43},
          doi = {10.1086/171833},
       adsurl = {https://scixplorer.org/abs/1992ApJ...398...43C}
}

@article{Dehnen_2017,
    author = {Dehnen, Walter},
    title = "{Towards time symmetric N-body integration}",
    journal = {\mnras},
    volume = {472},
    number = {1},
    pages = {1226-1238},
    year = {2017},
    month = {08},
    issn = {0035-8711},
    doi = {10.1093/mnras/stx1944},
    url = {https://doi.org/10.1093/mnras/stx1944},
    eprint = {https://academic.oup.com/mnras/article-pdf/472/1/1226/19730103/stx1944.pdf},
}

@article{Dimarco_2014,
    title={Numerical methods for kinetic equations},
    volume={23},
    DOI={10.1017/S0962492914000063}, journal={Acta Numerica},
    author={Dimarco, G. and Pareschi, L.}, year={2014},
    pages={369–520}}

@ARTICLE{Dolag_2026,
       author = {{Dolag}, Klaus and {Karademir}, Geray S. and {Tornatore}, Luca and {Ragagnin}, Antonio and {Valentini}, Milena and {B{\"o}ss}, Ludwig M. and {Groth}, Frederick and {Murante}, Giuseppe and {Borgani}, Stefano and {Springel}, Volker and et al.},
        title = "{The OpenGadget3 Code for Cosmological Simulations}",
      journal = {arXiv e-prints},
         year = 2026,
        month = sep,
          eid = {arXiv:2609.17658},
        pages = {arXiv:2609.17658},
          doi = {10.48550/arXiv.2609.17658},
archivePrefix = {arXiv},
       eprint = {2609.17658},
 primaryClass = {astro-ph.IM},
       adsurl = {https://ui.adsabs.harvard.edu/abs/2026arXiv260917658D}
}

@article{Dutton_2014,
   title={Cold dark matter haloes in the Planck era: evolution of structural parameters for Einasto and NFW profiles},
   volume={441},
   ISSN={0035-8711},
   url={http://dx.doi.org/10.1093/mnras/stu742},
   DOI={10.1093/mnras/stu742},
   number={4},
   journal={\mnras},
   publisher={Oxford University Press (OUP)},
   author={Dutton, Aaron A. and Macciò, Andrea V.},
   year={2014},
   month=may, pages={3359–3374} }

@article{Dvorkin_2014,
   title={Constraining dark matter-baryon scattering with linear cosmology},
   volume={89},
   ISSN={1550-2368},
   url={http://dx.doi.org/10.1103/PhysRevD.89.023519},
   DOI={10.1103/physrevd.89.023519},
   number={2},
   journal={\prd},
   publisher={American Physical Society (APS)},
   author={Dvorkin, Cora and Blum, Kfir and Kamionkowski, Marc},
   year={2014},
   month=jan }

@article{Ebisu_2022,
   title={Constraining self-interacting dark matter with dwarf spheroidal galaxies and high-resolution cosmological N-body simulations},
   volume={105},
   ISSN={2470-0029},
   url={http://dx.doi.org/10.1103/PhysRevD.105.023016},
   DOI={10.1103/physrevd.105.023016},
   number={2},
   journal={\prd},
   publisher={American Physical Society (APS)},
   author={Ebisu, Toshihiro and Ishiyama, Tomoaki and Hayashi, Kohei},
   year={2022},
   month={Jan}
}

@article{Enzi_2024,
    author = "Enzi, Wolfgang J. R. and Krawczyk, Coleman M. and Ballard, Daniel J. and Collett, Thomas E.",
    title = "{The overconcentrated dark halo in the strong lens SDSS J0946~+~1006 is a subhalo: evidence for self-interacting dark matter?}",
    eprint = "2411.08565",
    archivePrefix = "arXiv",
    primaryClass = "astro-ph.CO",
    doi = "10.1093/mnras/staf697",
    journal = {\mnras},
    volume = "540",
    number = "1",
    pages = "247--263",
    year = "2025"
}

@article{Fischer_2021a,
    author = {Fischer, Moritz S and Brüggen, Marcus and Schmidt-Hoberg, Kai and Dolag, Klaus and Kahlhoefer, Felix and Ragagnin, Antonio and Robertson, Andrew},
    title = "{N-body simulations of dark matter with frequent self-interactions}",
    journal = {\mnras},
    volume = {505},
    number = {1},
    pages = {851-868},
    year = {2021},
    month = {04},
    issn = {0035-8711},
    doi = {10.1093/mnras/stab1198},
    url = {https://doi.org/10.1093/mnras/stab1198},
    eprint = {https://academic.oup.com/mnras/article-pdf/505/1/851/38378730/stab1198.pdf},
}

@article{Fischer_2021b,
    author = {Fischer, Moritz S and Brüggen, Marcus and Schmidt-Hoberg, Kai and Dolag, Klaus and Ragagnin, Antonio and Robertson, Andrew},
    title = "{Unequal-mass mergers of dark matter haloes with rare and frequent self-interactions}",
    journal = {\mnras},
    volume = {510},
    number = {3},
    pages = {4080-4099},
    year = {2021},
    month = {12},
    issn = {0035-8711},
    doi = {10.1093/mnras/stab3544},
    url = {https://doi.org/10.1093/mnras/stab3544},
    eprint = {https://academic.oup.com/mnras/article-pdf/510/3/4080/42194749/stab3544.pdf},
}

@article{Fischer_2022,
    author = {Fischer, Moritz S and Brüggen, Marcus and Schmidt-Hoberg, Kai and Dolag, Klaus and Kahlhoefer, Felix and Ragagnin, Antonio and Robertson, Andrew},
    title = "{Cosmological simulations with rare and frequent dark matter self-interactions}",
    journal = {\mnras},
    volume = {516},
    number = {2},
    pages = {1923-1940},
    year = {2022},
    month = {08},
    issn = {0035-8711},
    doi = {10.1093/mnras/stac2207},
    url = {https://doi.org/10.1093/mnras/stac2207},
    eprint = {https://academic.oup.com/mnras/article-pdf/516/2/1923/45736085/stac2207.pdf},
}

@article{Fischer_2024b,
   title={Numerical challenges for energy conservation in N-body simulations of collapsing self-interacting dark matter halos},
   volume={689},
   ISSN={1432-0746},
   url={http://dx.doi.org/10.1051/0004-6361/202449849},
   DOI={10.1051/0004-6361/202449849},
   journal={\aap},
   publisher={EDP Sciences},
   author={Fischer, Moritz S. and Dolag, Klaus and Yu, Hai-Bo},
   year={2024},
   month=sep, pages={A300}
}

@ARTICLE{Fischer_2024c,
       author = {{Fischer}, Moritz S. and {Sagunski}, Laura},
        title = "{Dynamical friction from self-interacting dark matter}",
      journal = {\aap},
         year = 2024,
        month = oct,
       volume = {690},
          eid = {A299},
        pages = {A299},
          doi = {10.1051/0004-6361/202451304},
archivePrefix = {arXiv},
       eprint = {2405.19392},
 primaryClass = {astro-ph.CO},
       adsurl = {https://ui.adsabs.harvard.edu/abs/2024A&A...690A.299F}
}

@ARTICLE{Fischer_2025a,
       author = {{Fischer}, Moritz S. and {Dolag}, Klaus and {Garny}, Mathias and {Gluscevic}, Vera and {Groth}, Frederick and {Nadler}, Ethan O.},
        title = "{N-body simulations of dark matter─baryon interactions}",
      journal = {\aap},
         year = 2025,
        month = aug,
       volume = {700},
          eid = {A145},
        pages = {A145},
          doi = {10.1051/0004-6361/202554983},
archivePrefix = {arXiv},
       eprint = {2504.12393},
 primaryClass = {astro-ph.CO},
       adsurl = {https://ui.adsabs.harvard.edu/abs/2025A&A...700A.145F}
}

@ARTICLE{Fischer_2025b,
       author = {{Fischer}, Moritz S. and {Yu}, Hai-Bo and {Dolag}, Klaus},
        title = "{Accurately simulating core-collapse self-interacting dark matter halos}",
      journal = {\aap},
         year = 2025,
        month = nov,
       volume = {703},
          eid = {A234},
        pages = {A234},
          doi = {10.1051/0004-6361/202556189},
archivePrefix = {arXiv},
       eprint = {2506.06269},
 primaryClass = {astro-ph.CO},
       adsurl = {https://ui.adsabs.harvard.edu/abs/2025A&A...703A.234F}
}

@ARTICLE{Fischer_2026a,
       author = {{Fischer}, Moritz S. and {Yu}, Hai-Bo},
        title = "{The dark fate of ultra-faint dwarfs: Gravothermal collapse in action}",
      journal = {\aap},
         year = 2026,
        month = jul,
       volume = {711},
          eid = {A68},
        pages = {A68},
          doi = {10.1051/0004-6361/202659748},
archivePrefix = {arXiv},
       eprint = {2603.04508},
 primaryClass = {astro-ph.CO},
       adsurl = {https://ui.adsabs.harvard.edu/abs/2026A&A...711A..68F}
}

@ARTICLE{Fischer_2026b,
       author = {{Fischer}, Moritz S. and {Wiertel}, Marc and {Arido}, Cenanda and {Patil}, Yashraj and {Ragagnin}, Antonio and {Dolag}, Klaus and {Br{\"u}ggen}, Marcus and {Garny}, Mathias and {Robertson}, Andrew and {Schmidt-Hoberg}, Kai},
        title = "{The Cosmological Simulation Code OpenGadget3 - Implementation of Self-Interacting Dark Matter}",
      journal = {The Open Journal of Astrophysics},
         year = 2026,
        month = sep,
       volume = {9},
        pages = {71159},
          doi = {10.33232/001c.171159},
archivePrefix = {arXiv},
       eprint = {2603.10107},
 primaryClass = {astro-ph.IM},
       adsurl = {https://ui.adsabs.harvard.edu/abs/2026OJAp....971159F}
}

@article{Garrison-Kimmel_2013,
    author = "Garrison-Kimmel, Shea and Rocha, Miguel and Boylan-Kolchin, Michael and Bullock, James and Lally, Jaspreet",
    title = "{Can Feedback Solve the Too Big to Fail Problem?}",
    eprint = "1301.3137",
    archivePrefix = "arXiv",
    primaryClass = "astro-ph.CO",
    doi = "10.1093/mnras/stt984",
    journal = {\mnras},
    volume = "433",
    pages = "3539",
    year = "2013"
}

@ARTICLE{Gilman_2026a,
       author = {{Gilman}, D. and {Nierenberg}, A.~M. and {Gurian}, J. and {Paugnat}, H. and {Gannon}, C. and {Martinez}, M.~N. and {Treu}, T. and {Abazajian}, K.~N. and {Anguita}, T. and {Bennert}, V.~N. and et al.},
        title = "{JWST lensed quasar dark matter survey V: Hints of self-interacting dark matter from 29 quadruply imaged quasars}",
      journal = {arXiv e-prints},
         year = 2026,
        month = sep,
          eid = {arXiv:2609.35974},
        pages = {arXiv:2609.35974},
          doi = {10.48550/arXiv.2609.35974},
archivePrefix = {arXiv},
       eprint = {2609.35974},
 primaryClass = {astro-ph.CO},
       adsurl = {https://ui.adsabs.harvard.edu/abs/2026arXiv260935974G}
}

@ARTICLE{Gilman_2026b,
       author = {{Gilman}, Daniel and {Petraki}, Kalliopi and {Pignard}, Thomas and {Gurian}, James and {Kusenko}, Alex and {Nierenberg}, Anna},
        title = "{Hints of self-interacting dark matter from strong gravitational lensing}",
      journal = {arXiv e-prints},
         year = 2026,
        month = oct,
          eid = {arXiv:2610.40766},
        pages = {arXiv:2610.40766},
          doi = {10.48550/arXiv.2610.40766},
archivePrefix = {arXiv},
       eprint = {2610.40766},
 primaryClass = {astro-ph.CO},
       adsurl = {https://ui.adsabs.harvard.edu/abs/2026arXiv261040766G}
}

@ARTICLE{Gurian_2026,
       author = {{Gurian}, James and {Seher Gandhi}, Suroor and {Dalal}, Neal and {May}, Simon},
        title = "{Gravothermal Evolution of Self-Interacting Dark Matter Halos across Knudsen Numbers}",
      journal = {arXiv e-prints},
         year = 2026,
        month = oct,
          eid = {arXiv:2610.28745},
        pages = {arXiv:2610.28745},
          doi = {10.48550/arXiv.2610.28745},
archivePrefix = {arXiv},
       eprint = {2610.28745},
 primaryClass = {astro-ph.CO},
       adsurl = {https://ui.adsabs.harvard.edu/abs/2026arXiv261028745G}
}

@ARTICLE{Harris_2020,
  author  = {Harris, Charles R. and Millman, K. Jarrod and
            van der Walt, St\'{e}fan J and Gommers, Ralf and
            Virtanen, Pauli and Cournapeau, David and
            Wieser, Eric and Taylor, Julian and Berg, Sebastian and
            Smith, Nathaniel J. and Kern, Robert and Picus, Matti and
            Hoyer, Stephan and van Kerkwijk, Marten H. and
            Brett, Matthew and Haldane, Allan and
            Fern\'{a}ndez del R\'{i}o, Jaime and Wiebe, Mark and
            Peterson, Pearu and G\'{e}rard-Marchant, Pierre and
            Sheppard, Kevin and Reddy, Tyler and Weckesser, Warren and
            Abbasi, Hameer and Gohlke, Christoph and
            Oliphant, Travis E.},
  title   = {Array programming with {NumPy}},
  journal = {Nature},
  year    = {2020},
  volume  = {585},
  pages   = {357--362},
  doi     = {10.1038/s41586-020-2649-2}
}

@ARTICLE{Hunter_2007,
  author={J. D. {Hunter}},
  journal={Computing in Science   Engineering},
  title={Matplotlib: A 2D Graphics Environment},
  year={2007},
  volume={9},
  number={3},
  pages={90-95},
}

@article{Huo_2019,
    author = "Huo, Ran and Yu, Hai-Bo and Zhong, Yi-Ming",
    title = "{The Structure of Dissipative Dark Matter Halos}",
    eprint = "1912.06757",
    archivePrefix = "arXiv",
    primaryClass = "astro-ph.CO",
    doi = "10.1088/1475-7516/2020/06/051",
    journal = {\jcap},
    volume = "06",
    pages = "051",
    year = "2020"
}

@ARTICLE{Kahlhoefer_2014,
       author = {{Kahlhoefer}, Felix and {Schmidt-Hoberg}, Kai and {Frandsen}, Mads T. and {Sarkar}, Subir},
        title = "{Colliding clusters and dark matter self-interactions}",
      journal = {\mnras},
         year = 2014,
        month = jan,
       volume = {437},
       number = {3},
        pages = {2865-2881},
          doi = {10.1093/mnras/stt2097},
archivePrefix = {arXiv},
       eprint = {1308.3419},
 primaryClass = {astro-ph.CO},
       adsurl = {https://scixplorer.org/abs/2014MNRAS.437.2865K}
}

@ARTICLE{Karademir_2026,
       author = {{Karademir}, Geray S. and {Dolag}, Klaus},
        title = "{SPACE-Timers {\textemdash} A stack-based hierarchical timing system for C++}",
      journal = {Astronomy and Computing},
         year = 2026,
        month = jul,
       volume = {56},
          eid = {101122},
        pages = {101122},
          doi = {10.1016/j.ascom.2026.101122},
archivePrefix = {arXiv},
       eprint = {2603.01618},
 primaryClass = {astro-ph.IM},
       adsurl = {https://ui.adsabs.harvard.edu/abs/2026A&C....5601122K}
}

@ARTICLE{Klemmer_2026,
       author = {{Klemmer}, David and {Fischer}, Moritz S. and {Boddy}, Kimberly K. and {Kaplinghat}, Manoj and {Sagunski}, Laura},
        title = "{Survival of the most compact: the life and death of satellite halos in self-interacting dark matter}",
      journal = {arXiv e-prints},
         year = 2026,
        month = mar,
          eid = {arXiv:2603.19362},
        pages = {arXiv:2603.19362},
          doi = {10.48550/arXiv.2603.19362},
archivePrefix = {arXiv},
       eprint = {2603.19362},
 primaryClass = {astro-ph.GA},
       adsurl = {https://ui.adsabs.harvard.edu/abs/2026arXiv260319362K}
}

@ARTICLE{Kochanek_2000,
       author = {{Kochanek}, C.~S. and {White}, Martin},
        title = "{A Quantitative Study of Interacting Dark Matter in Halos}",
      journal = {\apj},
         year = "2000",
        month = "Nov",
       volume = {543},
       number = {2},
        pages = {514-520},
          doi = {10.1086/317149},
archivePrefix = {arXiv},
       eprint = {astro-ph/0003483},
 primaryClass = {astro-ph}
}

@ARTICLE{Koda_2011,
       author = {{Koda}, Jun and {Shapiro}, Paul R.},
        title = "{Gravothermal collapse of isolated self-interacting dark matter haloes: N-body simulation versus the fluid model}",
      journal = {\mnras},
         year = "2011",
        month = "Aug",
       volume = {415},
       number = {2},
        pages = {1125-1137},
          doi = {10.1111/j.1365-2966.2011.18684.x},
archivePrefix = {arXiv},
       eprint = {1101.3097},
 primaryClass = {astro-ph.CO},
       adsurl = {https://ui.adsabs.harvard.edu/abs/2011MNRAS.415.1125K}
}

@article{Kollmann_2025,
    author = "Kollmann, Kassidy E. and Nightingale, James W. and Lisanti, Mariangela and Robertson, Andrew and Slone, Oren",
    title = "{Using strong lensing to detect subhaloes with steep inner density profiles}",
    eprint = "2510.17956",
    archivePrefix = "arXiv",
    primaryClass = "astro-ph.CO",
    doi = "10.1093/mnras/stag066",
    journal = {\mnras},
    volume = "546",
    number = "4",
    pages = "stag066",
    year = "2026"
}

@article{Komatsu_2011,
	doi = {10.1088/0067-0049/192/2/18},
	url = {https://doi.org/10.1088/0067-0049/192/2/18},
	year = 2011,
	month = {jan},
	publisher = {American Astronomical Society},
	volume = {192},
	number = {2},
	pages = {18},
	author = {E. Komatsu and K. M. Smith and J. Dunkley and C. L. Bennett and B. Gold and G. Hinshaw and N. Jarosik and D. Larson and M. R. Nolta and L. Page and D. N. Spergel and M. Halpern and R. S. Hill and A. Kogut and M. Limon and S. S. Meyer and N. Odegard and G. S. Tucker and J. L. Weiland and E. Wollack and E. L. Wright},
	title = {{SEVEN}-{YEAR}
		                    $\less$i$\greater${WILKINSON} {MICROWAVE} {ANISOTROPY} {PROBE}$\less$/i$\greater$
		                    (
		                    $\less$i$\greater${WMAP}$\less$/i$\greater$
		                    ) {OBSERVATIONS}: {COSMOLOGICAL} {INTERPRETATION}},
	journal = {\apjs}
}

@ARTICLE{Kong_2026,
       author = {{Kong}, Demao and {Nadler}, Ethan O. and {Yu}, Hai-Bo},
        title = "{Strong-lensing perturber signatures in self-interacting dark matter simulations}",
      journal = {\prd},
         year = 2026,
        month = aug,
       volume = {114},
       number = {4},
          eid = {043011},
        pages = {043011},
          doi = {10.1103/xrqk-1w95},
archivePrefix = {arXiv},
       eprint = {2510.01491},
 primaryClass = {astro-ph.CO},
       adsurl = {https://scixplorer.org/abs/2026PhRvD.114d3011K}
}

@article{Li_2025,
    author = "Li, Shubo and others",
    title = "{The {\textquotedblleft}Little Dark Dot{\textquotedblright}: Evidence for Self-interacting Dark Matter in the Strong Lens SDSS J0946+1006?}",
    eprint = "2504.11800",
    archivePrefix = "arXiv",
    primaryClass = "astro-ph.GA",
    doi = "10.3847/1538-4357/ae1462",
    journal = {\apj},
    volume = "994",
    number = "2",
    pages = "201",
    year = "2025"
}

@ARTICLE{Li_2026,
       author = {{Li}, Shubo and {Fischer}, Moritz S. and {Jia}, Zixiang and {Jiang}, Fangzhou and {Li}, Ran and {Yu}, Hai-Bo},
        title = "{Testing the Isothermal Jeans Model for Self-interacting Dark Matter Halos in the Collapse Phase}",
      journal = {\apj},
         year = 2026,
        month = oct,
       volume = {1010},
       number = {2},
          eid = {197},
        pages = {197},
          doi = {10.3847/1538-4357/aea78a},
archivePrefix = {arXiv},
       eprint = {2603.01772},
 primaryClass = {astro-ph.CO},
       adsurl = {https://ui.adsabs.harvard.edu/abs/2026ApJ..1010..197L}
}

@ARTICLE{Lynden-Bell_1980,
       author = {{Lynden-Bell}, D. and {Eggleton}, P.~P.},
        title = "{On the consequences of the gravothermal catastrophe}",
      journal = {\mnras},
         year = 1980,
        month = may,
       volume = {191},
        pages = {483-498},
          doi = {10.1093/mnras/191.3.483},
       adsurl = {https://ui.adsabs.harvard.edu/abs/1980MNRAS.191..483L}
}

@ARTICLE{Mace_2024,
       author = {{Mace}, Charlie and {Zeng}, Zhichao Carton and {Peter}, Annika H.~G. and {Du}, Xiaolong and {Yang}, Shengqi and {Benson}, Andrew and {Vogelsberger}, Mark},
        title = "{Convergence tests of self-interacting dark matter simulations}",
      journal = {\prd},
         year = 2024,
        month = dec,
       volume = {110},
       number = {12},
          eid = {123024},
        pages = {123024},
          doi = {10.1103/PhysRevD.110.123024},
archivePrefix = {arXiv},
       eprint = {2402.01604},
 primaryClass = {astro-ph.GA},
       adsurl = {https://ui.adsabs.harvard.edu/abs/2024PhRvD.110l3024M}
}

@ARTICLE{Mace_2025,
       author = {{Mace}, Charlie and {Yang}, Shengqi and {Carton Zeng}, Zhichao and {Peter}, Annika H.~G. and {Du}, Xiaolong and {Benson}, Andrew},
        title = "{Calibrating the SIDM Gravothermal Catastrophe with N-body Simulations}",
      journal = {arXiv e-prints},
         year = 2025,
        month = apr,
          eid = {arXiv:2504.13004},
        pages = {arXiv:2504.13004},
          doi = {10.48550/arXiv.2504.13004},
archivePrefix = {arXiv},
       eprint = {2504.13004},
 primaryClass = {astro-ph.GA},
       adsurl = {https://ui.adsabs.harvard.edu/abs/2025arXiv250413004M}
}

@article{Minor_2020,
    author = "Minor, Quinn E. and Gad-Nasr, Sophia and Kaplinghat, Manoj and Vegetti, Simona",
    title = "{An unexpected high concentration for the dark substructure in the gravitational lens SDSSJ0946+1006}",
    eprint = "2011.10627",
    archivePrefix = "arXiv",
    primaryClass = "astro-ph.GA",
    doi = "10.1093/mnras/stab2247",
    journal = {\mnras},
    volume = "507",
    number = "2",
    pages = "1662--1683",
    year = "2021"
}

@article{Nadler_2023,
   title={A Self-interacting Dark Matter Solution to the Extreme Diversity of Low-mass Halo Properties},
   volume={958},
   ISSN={2041-8213},
   url={http://dx.doi.org/10.3847/2041-8213/ad0e09},
   DOI={10.3847/2041-8213/ad0e09},
   number={2},
   journal={\apjl},
   publisher={American Astronomical Society},
   author={Nadler, Ethan O. and Yang, Daneng and Yu, Hai-Bo},
   year={2023},
   month=nov, pages={L39} }

@ARTICLE{Nadler_2026,
       author = {{Nadler}, Ethan O. and {Rogers}, Keir K. and {Drlica-Wagner}, Alex},
        title = "{Dark Matter Constraints from Small-Scale Cosmic Structure}",
      journal = {arXiv e-prints},
         year = 2026,
        month = jul,
          eid = {arXiv:2607.28564},
        pages = {arXiv:2607.28564},
          doi = {10.48550/arXiv.2607.28564},
archivePrefix = {arXiv},
       eprint = {2607.28564},
 primaryClass = {astro-ph.CO},
       adsurl = {https://scixplorer.org/abs/2026arXiv260728564N}
}

@ARTICLE{Navarro_1996,
       author = {{Navarro}, Julio F. and {Frenk}, Carlos S. and {White}, Simon D.~M.},
        title = "{The Structure of Cold Dark Matter Halos}",
      journal = {\apj},
         year = 1996,
        month = may,
       volume = {462},
        pages = {563},
          doi = {10.1086/177173},
archivePrefix = {arXiv},
       eprint = {astro-ph/9508025},
 primaryClass = {astro-ph},
}

@article{Olson_2008,
title = {Gridless DSMC},
journal = {Journal of Computational Physics},
volume = {227},
number = {17},
pages = {8035-8064},
year = {2008},
issn = {0021-9991},
doi = {https://doi.org/10.1016/j.jcp.2008.04.038},
url = {https://www.sciencedirect.com/science/article/pii/S002199910800257X},
author = {Spencer E. Olson and Andrew J. Christlieb}
}

@ARTICLE{Palubski_2024,
       author = {{Palubski}, Igor and {Slone}, Oren and {Kaplinghat}, Manoj and {Lisanti}, Mariangela and {Jiang}, Fangzhou},
        title = "{Numerical challenges in modeling gravothermal collapse in Self-Interacting Dark Matter halos}",
      journal = {\jcap},
         year = 2024,
        month = sep,
       volume = {2024},
       number = {9},
          eid = {074},
        pages = {074},
          doi = {10.1088/1475-7516/2024/09/074},
archivePrefix = {arXiv},
       eprint = {2402.12452},
 primaryClass = {astro-ph.CO},
       adsurl = {https://ui.adsabs.harvard.edu/abs/2024JCAP...09..074P}
}

@article{Patil_2026,
	author = {{Patil}, Yashraj and {Fischer}, Moritz S.},
	title = {To collapse or not to collapse: Halo evolution with self-interacting dark matter mass segregation},
	DOI= "10.1051/0004-6361/202556178",
	url= "https://doi.org/10.1051/0004-6361/202556178",
	journal = {\aap},
	year = 2026,
	volume = 715,
	pages = "A70",
}

@INPROCEEDINGS{Ragagnin_2016,
       author = {{Ragagnin}, Antonio and {Tchipev}, Nikola and {Bader}, Michael and {Dolag}, Klaus and {Hammer}, Nicolay J.},
        title = "{Exploiting the Space Filling Curve Ordering of Particles in the Neighbour Search of Gadget3}",
    booktitle = {Advances in Parallel Computing},
         year = 2016,
        month = may,
        pages = {411-420},
          doi = {10.3233/978-1-61499-621-7-411},
archivePrefix = {arXiv},
       eprint = {1810.09898},
 primaryClass = {astro-ph.IM},
       adsurl = {https://ui.adsabs.harvard.edu/abs/2016pcre.conf..411R}
}

@ARTICLE{Ragagnin_2024,
       author = {{Ragagnin}, A. and {Meneghetti}, M. and {Calura}, F. and {Despali}, G. and {Dolag}, K. and {Fischer}, M.~S. and {Giocoli}, C. and {Moscardini}, L.},
        title = "{Dianoga SIDM: Galaxy cluster self-interacting dark matter simulations}",
      journal = {\aap},
         year = 2024,
        month = jul,
       volume = {687},
          eid = {A270},
        pages = {A270},
          doi = {10.1051/0004-6361/202449872},
archivePrefix = {arXiv},
       eprint = {2404.01383},
 primaryClass = {astro-ph.CO},
       adsurl = {https://ui.adsabs.harvard.edu/abs/2024A&A...687A.270R}
}

@article{Robles_2017,
   title={SIDM on fire: hydrodynamical self-interacting dark matter simulations of low-mass dwarf galaxies},
   volume={472},
   ISSN={1365-2966},
   url={http://dx.doi.org/10.1093/mnras/stx2253},
   DOI={10.1093/mnras/stx2253},
   number={3},
   journal={\mnras},
   publisher={Oxford University Press (OUP)},
   author={Robles, Victor H. and Bullock, James S. and Elbert, Oliver D. and Fitts, Alex and González-Samaniego, Alejandro and Boylan-Kolchin, Michael and Hopkins, Philip F. and Faucher-Giguère, Claude-André and Kereš, Dušan and Hayward, Christopher C.},
   year={2017},
   month=sep, pages={2945–2954} }

@ARTICLE{Robertson_2017a,
       author = {{Robertson}, Andrew and {Massey}, Richard and {Eke}, Vincent},
        title = "{What does the Bullet Cluster tell us about self-interacting dark matter?}",
      journal = {\mnras},
         year = "2017",
        month = "Feb",
       volume = {465},
       number = {1},
        pages = {569-587},
          doi = {10.1093/mnras/stw2670},
archivePrefix = {arXiv},
       eprint = {1605.04307},
 primaryClass = {astro-ph.CO},
       adsurl = {https://ui.adsabs.harvard.edu/abs/2017MNRAS.465..569R}
}

@ARTICLE{Rosenlyst_2026,
       author = {{Rosenlyst}, Martin and {Tran}, Vinh and {Zier}, Oliver and {Shen}, Xuejian and {Vogelsberger}, Mark and {Brinckmann}, Thejs and {Frandsen}, Mads T. and {Zavala}, Jes{\'u}s and {Holt Dihver Christensen}, Frederik and {Theiss Kristensen}, Mikkel},
        title = "{Self-Interacting Dark Matter: From Particle Physics Theories to Structure Formation Simulations}",
      journal = {arXiv e-prints},
         year = 2026,
        month = oct,
          eid = {arXiv:2610.16968},
        pages = {arXiv:2610.16968},
          doi = {10.48550/arXiv.2610.16968},
archivePrefix = {arXiv},
       eprint = {2610.16968},
 primaryClass = {astro-ph.CO},
       adsurl = {https://ui.adsabs.harvard.edu/abs/2026arXiv261016968R}
}

@ARTICLE{Schmidt_2026,
       author = {{Schmidt}, Ludwig D. and {Fischer}, Moritz S. and {Garny}, Mathias},
        title = "{Cooling, conduction, and compact objects: Gravothermal evolution of dissipative self-interacting dark matter halos}",
      journal = {\aap},
         year = 2026,
        month = oct,
       volume = {714},
          eid = {A469},
        pages = {A469},
          doi = {10.1051/0004-6361/202661917},
archivePrefix = {arXiv},
       eprint = {2606.19428},
 primaryClass = {astro-ph.CO},
       adsurl = {https://ui.adsabs.harvard.edu/abs/2026A&A...714A.469S}
}

@article{Spergel_2000,
   title={Observational Evidence for Self-Interacting Cold Dark Matter},
   volume={84},
   ISSN={1079-7114},
   url={http://dx.doi.org/10.1103/PhysRevLett.84.3760},
   DOI={10.1103/physrevlett.84.3760},
   number={17},
   journal={\prl},
   publisher={American Physical Society (APS)},
   author={Spergel, David N. and Steinhardt, Paul J.},
   year={2000},
   month={Apr},
   pages={3760--3763}
}

@ARTICLE{Springel_2005,
       author = {{Springel}, Volker},
        title = "{The cosmological simulation code GADGET-2}",
      journal = {\mnras},
         year = "2005",
        month = "Dec",
       volume = {364},
       number = {4},
        pages = {1105-1134},
          doi = {10.1111/j.1365-2966.2005.09655.x},
archivePrefix = {arXiv},
       eprint = {astro-ph/0505010},
 primaryClass = {astro-ph},
       adsurl = {https://ui.adsabs.harvard.edu/abs/2005MNRAS.364.1105S}
}

@MISC{Springel_2015,
       author = {{Springel}, Volker},
        title = "{N-GenIC: Cosmological structure initial conditions}",
         year = 2015,
        month = feb,
          eid = {ascl:1502.003},
        pages = {ascl:1502.003},
archivePrefix = {ascl},
       eprint = {1502.003},
       adsurl = {https://ui.adsabs.harvard.edu/abs/2015ascl.soft02003S}
}

@article{Tajalli_2025,
    author = "Tajalli, Maryam and Vegetti, Simona and O'Riordan, Conor M. and White, Simon D. M. and Fassnacht, Christopher D. and Powell, Devon M. and McKean, J. P. and Despali, Giulia",
    title = "{SHARP{\textendash}IX. The dense, low-mass perturbers in B1938+666 and J0946+1006: implications for cold and self-interacting dark matter}",
    eprint = "2505.07944",
    archivePrefix = "arXiv",
    primaryClass = "astro-ph.CO",
    doi = "10.1093/mnras/staf1357",
    journal = {\mnras},
    volume = "543",
    number = "1",
    pages = "540--564",
    year = "2025"
}

@ARTICLE{Tulin_2018,
       author = {{Tulin}, Sean and {Yu}, Hai-Bo},
        title = "{Dark matter self-interactions and small scale structure}",
      journal = {\physrep},
         year = "2018",
        month = "Feb",
       volume = {730},
        pages = {1-57},
          doi = {10.1016/j.physrep.2017.11.004},
archivePrefix = {arXiv},
       eprint = {1705.02358},
 primaryClass = {hep-ph},
       adsurl = {https://ui.adsabs.harvard.edu/abs/2018PhR...730....1T}
}

@article{Valdarnini_2024,
    author = "Valdarnini, R.",
    title = "{An N-body/hydrodynamical simulation study of the merging cluster El Gordo: A compelling case for self-interacting dark matter?}",
    eprint = "2309.10374",
    archivePrefix = "arXiv",
    primaryClass = "astro-ph.CO",
    doi = "10.1051/0004-6361/202348000",
    journal = {\aap},
    volume = "684",
    pages = "A102",
    year = "2024"
}

@article{Vegetti_2026,
   title={A possible challenge for cold and warm dark matter},
   ISSN={2397-3366},
   url={http://dx.doi.org/10.1038/s41550-025-02746-w},
   DOI={10.1038/s41550-025-02746-w},
   journal={Nature Astronomy},
   publisher={Springer Science and Business Media LLC},
   author={Vegetti, Simona and White, Simon D. M. and McKean, John P. and Powell, Devon M. and Spingola, Cristiana and Massari, Davide and Despali, Giulia and Fassnacht, Christopher D.},
   year={2026},
   month=jan
}

@article{Yang_2020,
   title={Self-Interacting Dark Matter and the Origin of Ultradiffuse Galaxies NGC1052-DF2 and -DF4},
   volume={125},
   ISSN={1079-7114},
   url={http://dx.doi.org/10.1103/PhysRevLett.125.111105},
   DOI={10.1103/physrevlett.125.111105},
   number={11},
   journal={\prl},
   publisher={American Physical Society (APS)},
   author={Yang, Daneng and Yu, Hai-Bo and An, Haipeng},
   year={2020},
   month={Sep}
}

@article{Yang_2021,
    author = "Yang, Daneng and Yu, Hai-Bo",
    title = "{Self-interacting dark matter and small-scale gravitational lenses in galaxy clusters}",
    eprint = "2102.02375",
    archivePrefix = "arXiv",
    primaryClass = "astro-ph.GA",
    doi = "10.1103/PhysRevD.104.103031",
    journal = {\prd},
    volume = "104",
    number = "10",
    pages = "103031",
    year = "2021"
}

@ARTICLE{Yu_2026,
       author = {{Yu}, Hai-Bo},
        title = "{Core-Collapsed SIDM Halos as the Common Origin of Dense Perturbers in Lenses, Streams, and Satellites}",
      journal = {\prl},
         year = 2026,
        month = apr,
       volume = {136},
       number = {14},
          eid = {141001},
        pages = {141001},
          doi = {10.1103/txxx-97ln},
archivePrefix = {arXiv},
       eprint = {2510.11006},
 primaryClass = {astro-ph.GA},
       adsurl = {https://ui.adsabs.harvard.edu/abs/2026PhRvL.136n1001Y}
}

@ARTICLE{Zeng_2022,
       author = {{Zeng}, Zhichao Carton and {Peter}, Annika H.~G. and {Du}, Xiaolong and {Benson}, Andrew and {Kim}, Stacy and {Jiang}, Fangzhou and {Cyr-Racine}, Francis-Yan and {Vogelsberger}, Mark},
        title = "{Core-collapse, evaporation, and tidal effects: the life story of a self-interacting dark matter subhalo}",
      journal = {\mnras},
         year = 2022,
        month = jul,
       volume = {513},
       number = {4},
        pages = {4845-4868},
          doi = {10.1093/mnras/stac1094},
archivePrefix = {arXiv},
       eprint = {2110.00259},
 primaryClass = {astro-ph.CO},
       adsurl = {https://ui.adsabs.harvard.edu/abs/2022MNRAS.513.4845Z}
}

@ARTICLE{Zeng_2026,
       author = {{Zeng}, Zhichao Carton and {Peter}, Annika H.~G. and {Strigari}, Louis E. and {Benson}, Andrew and {Du}, Xiaolong and {Mace}, Charlie},
        title = "{Feeding the Void: Co-evolution of the SIDM-seeded Black Hole and the dark halo after core-collapse}",
      journal = {arXiv e-prints},
         year = 2026,
        month = sep,
          eid = {arXiv:2609.22758},
        pages = {arXiv:2609.22758},
archivePrefix = {arXiv},
       eprint = {2609.22758},
 primaryClass = {astro-ph.CO},
       adsurl = {https://ui.adsabs.harvard.edu/abs/2026arXiv260922758C}
}

@article{Zhang_2025,
doi = {10.3847/2041-8213/ada02b},
url = {https://dx.doi.org/10.3847/2041-8213/ada02b},
year = {2025},
month = {jan},
publisher = {The American Astronomical Society},
volume = {978},
number = {2},
pages = {L23},
author = {Zhang, Xingyu and Yu, Hai-Bo and Yang, Daneng and Nadler, Ethan O.},
title = {The GD-1 Stellar Stream Perturber as a Core-collapsed Self-interacting Dark Matter Halo},
journal = {\apjl}
}

@article{Zhong_2023,
   title={The impact of baryonic potentials on the gravothermal evolution of self-interacting dark matter haloes},
   volume={526},
   ISSN={1365-2966},
   url={http://dx.doi.org/10.1093/mnras/stad2765},
   DOI={10.1093/mnras/stad2765},
   number={1},
   journal={\mnras},
   publisher={Oxford University Press (OUP)},
   author={Zhong, Yi-Ming and Yang, Daneng and Yu, Hai-Bo},
   year={2023},
   month=sep, pages={758–770} }

@ARTICLE{Zier_2026,
       author = {{Zier}, Oliver and {Shen}, Xuejian and {Tran}, Vinh and {Rosenlyst}, Martin and {Vogelsberger}, Mark and {Liu}, Rongrong},
        title = "{A Novel Implementation of Self-Interacting Dark Matter in AREPO}",
      journal = {arXiv e-prints},
         year = 2026,
        month = jul,
          eid = {arXiv:2607.05504},
        pages = {arXiv:2607.05504},
          doi = {10.48550/arXiv.2607.05504},
archivePrefix = {arXiv},
       eprint = {2607.05504},
 primaryClass = {astro-ph.CO},
       adsurl = {https://ui.adsabs.harvard.edu/abs/2026arXiv260705504Z}
}

\end{document}